\documentclass[prd,preprint,preprintnumbers,nofootinbib,eqsecnum,superscriptaddress]{revtex4}

 \usepackage[dvips,final]{graphicx}
  \usepackage{amssymb}
   \usepackage{amsmath}
    \usepackage{amsfonts}
     \usepackage{epsfig}
      \usepackage{bm}

\usepackage{mathpazo}

\usepackage{multirow}
\usepackage{ctable}
\usepackage{booktabs}
\usepackage{array}
\usepackage{tabularx}
\usepackage{xcolor}
\usepackage{pstricks}
\definecolor{fred}{rgb}{0.90053, 0.00369, 0.00159}  

\begin{document}

\title{Independent quark/antiquark fragmentation
to massive particles\\ in proton-proton collisions}

\author{Rafa{\l} Maciu{\l}a}
\email{rafal.maciula@ifj.edu.pl} \affiliation{Institute of Nuclear
Physics, Polish Academy of Sciences, Radzikowskiego 152, PL-31-342 Krak{\'o}w, Poland}

\author{Antoni Szczurek\footnote{also at University of Rzesz\'ow, PL-35-959 Rzesz\'ow, Poland}}
\email{antoni.szczurek@ifj.edu.pl} \affiliation{Institute of Nuclear
Physics, Polish Academy of Sciences, Radzikowskiego 152, PL-31-342 Krak{\'o}w, Poland}

\begin{abstract}
We critically discuss fragmentation of quark or antiquark to massive particles
(mesons or baryons) in proton-proton collisions.
Both heavy and light quark/antiquark fragmentations are discussed 
using universal $z$-dependent fragmentation functions.
Different scenarios how to define the $z$ variable are considered: 
as a fraction of energy, momentum or light-cone momentum 
of the parent quark/antiquark.
Also a choice of the direction of motion of hadron 
with respect to the parent parton must be made in the simplest approach.
Energy and flavour violation is discussed for the region
of small $p_t$ and/or center-of-mass $y \sim$ 0. 
Results of different approaches are compared.
We show that at the LHC energies all schemes become consistent for $D$-meson
transverse momenta larger than $2$ GeV.
Relations to results from the literature are made.
We present some examples for production of D mesons from light 
and charm quarks/antiquarks.
Emission with respect to the direction of motion lowers the cross
section of $D$ mesons at larger rapidities compared to the traditional
approach ($y_D = y_{q/\bar q}$).
As illustrated here the effect of using different prescriptions is
particularly large at low energies (fixed target experiments).
\end{abstract}

\maketitle

\section{Introduction}

The concept of using fragmentation functions was routinely used for 
production of mesons (or baryons) in $e^+ e^-$ collisions \cite{Peterson:1982ak,Cacciari:2005uk,Kniehl:2006mw,Kneesch:2007ey}.
A similar approach is commonly used in proton-proton collisions.
Both light $m_{H} <$ 0.5 GeV and heavy $m_{H} >$ 1.5 GeV hadrons ($H$) production were
considered within this framework. The $D$ or $B$ mesons (see \textit{e.g.} Refs.~\cite{Maciula:2013wg,Jung:2011yt}) or $\Lambda_c$ baryons \cite{Maciula:2018iuh} are good examples.
For high-energy $e^+ e^-$ collisions also quark/antiquark jets have rather large energies/momenta.
In proton-proton collisions only a part (often rather small) 
of the total energy goes to quark/antiquark production.
Often quark or antiquark has a small energy $E_{q} < m_H$, smaller than
the mass of the heavy object $H$. What to do then within a picture
of independent quark/antiquark fragmentation?

One usually assumes that $D$ mesons are produced from fragmentation
of $c$ quark or $\bar c$ antiquark. Peterson fragmentation functions 
are used usually in this context \cite{Peterson:1982ak}.
The LHCb observed asymmetries in production of $D^+/D^-$
\cite{LHCb:2012fb} and very recently also in $D_s^+/D_s^-$
\cite{Aaij:2018afd}. In Ref.~\cite{Maciula:2017wov} these asymmetries were explained in terms of
subleading fragmentation of light quark/antiquark to D mesons.
Such an approach involves using corresponding fragmentation functions
which are poorly known so far. In the mentioned references we suggested
how to limit them by the LHCb data for $D$-meson asymmetries.

In the present studies we are interested rather in small transverse momenta
of $D$ mesons. Even there the independent parton fragmentation picture
is applied for $c / \bar c \to D$ fragmentation.
It is usually assumed, somewhat arbitrarily, that $y_D = y_c \equiv y$
and for each $y$ (separately) convolution in transverse momentum is done. 
It is not checked in this context whether $E_D$ is larger than
the mass of the $D$ meson ($m_D$) and smaller than energy of the parent
quark/antiquark.
Another option is to assume that $D$ meson is emitted in the same
direction as the parent $c$ or $\bar c$.
For light quark/antiquark fragmentations the different approximations
may lead to different results. What is applicability of the independent
parton fragmentation picture?

In the present paper we wish to show how the approximation used change
the results for distributions of heavy hadrons and discuss some conceptual problems of some approaches
in different corners of the phase space. We wish to concentrate especially on small transverse momenta and forward/backward rapidities of hadrons as well as on low c.m.s. collision energies $\sqrt{s}$. These kinematical regimes are of the special importance \textit{e.g.}
for studies of high-energy prompt neutrino flux at IceCube and for heavy meson predictions devoted to
low energy experiments, like planned SHiP experiment. There, straight applications of the standard approximations introduced originally for massless particles and devoted rather to high energies and larger transverse momenta, where masses of both, parton and hadron can be neglected, seem to be too naive.

\section{Parton level calculations}

As already mentioned the main goal of this study is to discuss
fragmentation of light $q$ and heavy $Q$ quarks to heavy objects $H$ 
(heavy mesons or baryons).  
For clarity, we limit the following studies to the case of production of
$D_{s}^{\pm}$ meson where we have two fragmentation components: standard $c/\bar{c} \to D_{s}^{\pm}$ and unfavoured (subleading) $s/\bar{s} \to D_{s}^{\pm}$. Within this scenario, in the first step we need to calculate the parton-level cross sections for charm and strange quark/antiquark production.   

The cross section for $c\bar{c}$-pair production at high energies is dominated by the gluon-gluon fusion. This is also true at lower energies as long as one considers small transverse momenta and rather midrapidity regions, where the $q\bar q$-annihilation component still remains negligible. In the numerical calculations here, we follow the $k_{t}$-factorization approach where both incident gluons are off-mass shell and their emission is encoded in the so-called unintegrated (transverse momentum dependent) parton distribution functions (uPDFs). The transverse momenta (virtualities) of both partons entering the hard process are taken into account and the sum of transverse momenta of the final $c$ and $\bar c$ no longer cancels. Then the differential cross section at the tree-level for the $c \bar c$-pair production reads:
\begin{eqnarray}\label{LO_kt-factorization} 
\frac{d \sigma(p p \to c \bar c \, X)}{d y_1 d y_2 d^2p_{1,t} d^2p_{2,t}} &=&
\int \frac{d^2 k_{1,t}}{\pi} \frac{d^2 k_{2,t}}{\pi}
\frac{1}{16 \pi^2 (x_1 x_2 s)^2} \; \overline{ | {\cal M}^{\mathrm{off-shell}}_{g^* g^* \to c \bar c} |^2}
 \\  
&& \times  \; \delta^{2} \left( \vec{k}_{1,t} + \vec{k}_{2,t} 
                 - \vec{p}_{1,t} - \vec{p}_{2,t} \right) \;
{\cal F}_g(x_1,k_{1,t}^2) \; {\cal F}_g(x_2,k_{2,t}^2) \; \nonumber ,   
\end{eqnarray}
where ${\cal F}_g(x_1,k_{1,t}^2)$ and ${\cal F}_g(x_2,k_{2,t}^2)$
are the gluon uPDFs for both colliding hadrons and ${\cal M}^{\mathrm{off-shell}}_{g^* g^* \to c \bar c}$ is the off-shell matrix element for the hard $g g \to c \bar c$ subprocess. More details of the calculations can be found in our previous papers \cite{Maciula:2013wg,Maciula:2018iuh}. Here we use the Kimber-Martin-Ryskin (KMR) \cite{Watt:2003mx} gluon uPDF calculated from CTEQ6 \cite{Pumplin:2002vw} collinear PDFs. 

The cross section for s-quark production at low and high energies is dominated by the two subprocesses $g s \to g s$ and $s g \to s g$ (the same is true for $\bar{s}$-antiquark). In this case, the calculations are done in the leading-order (LO) collinear factorization approach with on-shell initial state partons and with a special
treatment of minijets at low transverse momenta, as adopted \textit{e.g.} in \textsc{Pythia}, 
by multiplying standard cross section by a somewhat arbitrary suppression factor \cite{Sjostrand:2014zea}. The cross section reads then
\begin{eqnarray}
\frac{d \sigma}{d y_1 d y_2 d^2p_t} &=& \frac{1}{16 \pi^2 {\hat s}^2}
[ x_1 g(x_1,\mu^2) \; x_2 s(x_2,\mu^2) \;
\overline{|{\cal M}_{gs\to gs}|^2} \; \\ \nonumber && + \; x_1 s(x_1,\mu^2) \; x_2 g(x_2,\mu^2) \;
\overline{|{\cal M}_{sg\to sg}|^2} ] \times F_{sup}(p_{t})  \; ,
\label{LO_collinear}
\end{eqnarray}
where $g(x,\mu^2)$ and $s(x_2,\mu^2)$ are the familiar (collinear) gluon and s-quark PDFs and
\begin{equation}
F_{sup}(p_t) = \frac{p_t^4}{((p_{t}^{0})^{2} + p_t^2)^2} \; .
\label{suppression_factor}
\end{equation}
Within this framework the cross section of course strongly depends on the free parameter $p_{t}^{0}$ which could be, in principle, fitted to low energy charm experimental data \cite{Maciula:2017wov}. Here, we use rather conservative value $p_{t}^{0} = 2$ GeV. 

\section{Quark to hadron fragmentation}

The transition from quarks and gluons to hadrons, called hadronization or parton fragmentation, can be so far approached only through phenomenological models. In principle, in the case of multi-particle final states the Lund string model \cite{Andersson:1983ia} and the cluster
fragmentation model \cite{Webber:1983if} are often used. However, following non-Monte-Carlo methods and considering fragmentation of not a complex parton system but of a single (separated) parton one usually follows independent parton fragmentation functions (FF) technique.

For instance, standard theoretical studies of inclusive open charm meson production at the LHC based on next-to-leading order (NLO) collinear approach within the FONLL scheme \cite{Cacciari} as well as on the $k_{t}$-factorization \cite{Maciula:2013wg} are usually done with the help of the scale-independent
FFs. In turn, in Ref.~\cite{Kniehl2012} the calculation was done according to the GM-VFNS NLO collinear scheme together with the several scale-dependent FFs of a parton (gluon, $u,d,s,\bar u, \bar d, \bar s, c, \bar c$) to $D$ mesons proposed by Kniehl et al. \cite{Kniehl:2006mw}, that undergo DGLAP evolution equations. Within this framework an important contribution to inclusive production of $D$ mesons comes from gluon fragmentation (see also Ref.~\cite{Kniehl:2005ej}). Similar calculation were done recently also in the
$k_t$-factorization approach with parton Reggeization hypothesis \cite{Nefedov:2014qea}. Here, we follow the framework with the scale-independent FFs and do not
consider effects of their evolution since our main goal is to discuss rather a basics concepts of the framework without a special emphasis on the form of
parametrizations of fragmentation functions.

\subsection{Standard approach}

According to the standard DGLAP-based formalism for the fragmentation, the inclusive distributions of heavy hadrons $H = D, B$ can be obtained through a convolution of inclusive distributions of heavy quarks/antiquarks $Q$ and $Q \to H$ fragmentation functions:
\begin{equation}
\frac{d \sigma(pp \rightarrow H \overline{H} \; X)}{d y_H d^2 p_{t,H}} \approx
\int_0^1 \frac{dz}{z^2} D_{Q \to H}(z)
\frac{d \sigma(pp \rightarrow Q \overline{Q} X)}{d y_Q d^2 p_{t,Q}}
\Bigg\vert_{y_Q = y_H \atop p_{t,Q} = p_{t,H}/z} \;,
\label{Q_to_h}
\end{equation}
where $p_{t,Q} = \frac{p_{t,H}}{z}$ and $z$ is the fraction of
longitudinal momentum of heavy quark $Q$ carried by a heavy hadron $H$.
Here the typical approximation is done that $y_{Q}$ is
unchanged in the fragmentation process, i.e. $y_H = y_Q$.
This commonly accepted and frequently used method was originally
proposed for light partons. It can be safely used only when both, mass of the parton and mass of the 
hadron can be neglected \cite{Maciula:2015kea}. In principle, this approximation may not be valid for the case of heavy and even light parton fragmentation to heavy object, especially, at lower energies or/and considering regions of small transverse momenta.

So far, applicability of this method for massive particles, to the best of our knowledge, was not discussed in the literature.
In many phenomenological studies of heavy meson production based on the independent parton fragmentation picture this approximation was applied \textit{a priori}. However, it is obvious that working with massive particles this approach may break down at small transverse momenta of a hadron, when approaching $p_{T} \sim m_H$ region. In this regime one could expect a violation of "energy conservation"\footnote{In the independent parton fragmentation picture one is not giving description of the hadronization of the parton system as a whole, so the
energy conservation has a special interpretation. Therefore it is written here in quotes.} and events with hadrons that have larger energies than the energy of the parent parton can frequently appear. In some corners of the phase space the $E_{H} < E_{q}$ relation may be broken very strongly.

As long as one is considering c.m.s. midrapidities and/or large c.m.s. collision energies this mass effect shall be rather negligible, especially, when a low transverse momentum cut is applied. However, the situation may dramatically change when going to lower energies.
In fact, this effect may become important even at larger energies, when
discussing forward (or far-forward) production. Therefore, we expect
this standard approach not to be valid \textit{e.g.} for studies of high-energy prompt neutrino flux at IceCube and for heavy meson predictions devoted to
low energy experiments, like planned SHiP experiment \cite{SHIP}.

In this context we wish to propose and discuss other prescriptions that could be an useful alternative in phenomenological studies
of heavy flavour production in different kinematical regimes.
 
\subsection{Emission in the same direction}

In contrast to the standard approach, here we follow a different idea and assume that the hadron $H$ is emitted in the direction of parent
quark/antiquark $q$, i.e. $\eta_H = \eta_q$ (the same pseudorapidities or
polar angles). Within this approach still different options for $z$-scaling come into game:
\begin{itemize}
\item $p_H = z p_q$ (momentum scaling),
\item $E_H = z E_q$ (energy scaling),
\item $p_H^+ = z p_q^+$ (light-cone scaling) where $p^{+} = E+p$.
\end{itemize}
In the case of energy scaling approach, in general, $z E_q$ can be smaller
than $m_H$ which is, at least naively, in conflict with 
``energy conservation'' in the parton-to-hadron process
if we take the parton as the only reservoir of energy (idependent
parton fragmentation). Thus, within this choice of scaling we include extra condition $z E_q > m_H$.
In other cases this condition is satisfied automatically by the definition of $z$.
We also include the condition $E_{D} \leq E_{q}$ which is strongly broken in the standard fragmentation framework with constant rapidity and which is present by definition only in the case of energy scaling.
The three proposed prescriptions reproduce the standard approach in the limit: $m_{q}, m_{D} \to 0$.

\begin{figure}[!h]
\begin{minipage}{0.47\textwidth}
  \centerline{\includegraphics[width=1.0\textwidth]{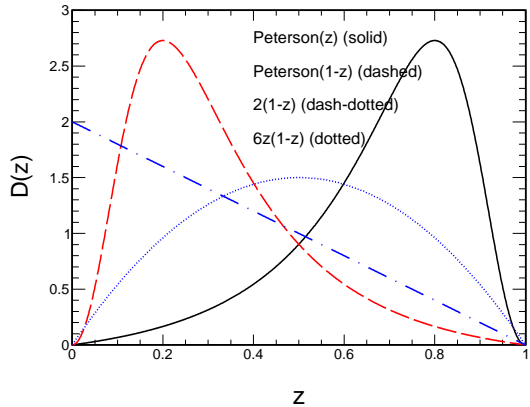}}
\end{minipage}
\caption{\small
Fragmentation functions used in the present analysis. Here $\int D(z) \; dz = 1$.
}
\label{fig:frag_functions}
\end{figure}

So far our considerations were rather general. In the present analysis we wish to consider two cases: light-to-heavy and heavy-to-heavy fragmentation.
Thus, as an example, we will show our predictions for $D_{s}$ meson production discussed recently in the context of
the LHCb production asymmetry \cite{Goncalves:2018zzf}, taking into account the standard $c/\bar c \to D_s^{\pm}$ and unfavoured
(subleading) $s / {\bar s} \to D_s^{\mp}$ fragmentation mechanisms. In such a calculation we need corresponding fragmentation functions.
For $c/\bar c \to D_s^{\pm}$ fragmentation we take traditional Peterson fragmentation function with $\varepsilon$ = 0.05.
In contrast to the standard mechanism, the fragmentation function for
$s/{\bar s} \to D_s^{\mp}$ transition is completely unknown which makes
the situation more difficult. For illustration we shall take therefore 
a few functional forms for the corresponding fragmentation functions:
\begin{itemize}
\item $D(z) = P \cdot \mathrm{Peterson}(1-z)$ (called reversed Peterson),
\item $D(z) = P \cdot 2 (1-z)$ (called triangle),
\item $D(z) = P \cdot 6z(1-z)$ (called hiperbolic).
\end{itemize}
The transition probability $P = P_{s \to D_s}$ can be treated as a free parameter and needs to be extracted
from experimental data. First attempt was done very recently in Ref.~\cite{Goncalves:2018zzf}, where $D^{+}_{s}/D^{-}_{s}$ production asymmetry was studied. In Fig.~\ref{fig:frag_functions} we illustrate
the shapes of the fragmentation functions used in the present analysis.

For light-to-heavy fragmentation one has to carefully check
the energy available for $s /\bar s \to D_s^{\mp}$ transition.
Of course a minimal condition is: $E_{D_{s}} > m_{D_{s}}$, i.e. there must be energy
available for the parton to produce the heavy meson. This means that
such $s$ or $\bar s$ that does not fullfill the energy condition
fragment rather to lighter mesons containing $s/{\bar s}$ such as 
$K, \eta, \phi$. 
The energy condition applies also to heavy-to-heavy fragmentation
but there are no much lighter mesons/baryons in this case. For $D_s$ these are
$D^{\pm}$ and $D^0/{\bar D}^0$ that have almost the same masses
so cannot be produced either. On the other hand in strong processes
flavour is conserved, so the damping of the D meson production rates
caused by the "energy conservation" needs a compensation by other mechanisms.
It is known that at sufficiently small scales (small invariant masses) the $c\bar{c}$-pair may likely hadronize
into quarkonia bound states. This may explain a part of the "missing" charm strength in the $c \to D$ haronization but certainly is not giving a final solution. Within this problematic region of phase space effects of parton recombination or other non-perturbative effects may prove to be crucial in this context, however, this requires further studies.

\section{Numerical results}

\begin{figure}[!h]
\begin{minipage}{0.47\textwidth}
  \centerline{\includegraphics[width=1.0\textwidth]{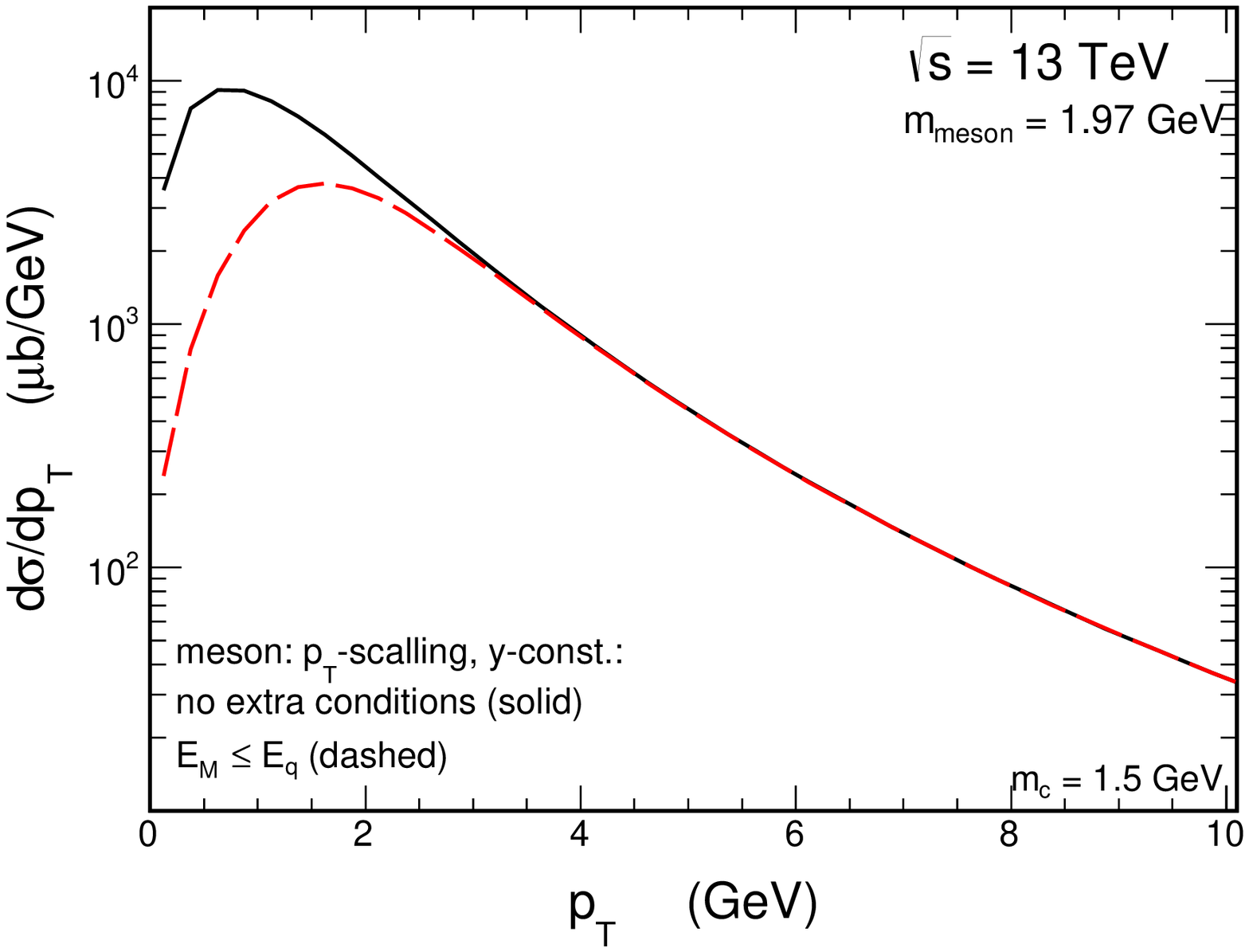}}
\end{minipage}
\begin{minipage}{0.47\textwidth}
  \centerline{\includegraphics[width=1.0\textwidth]{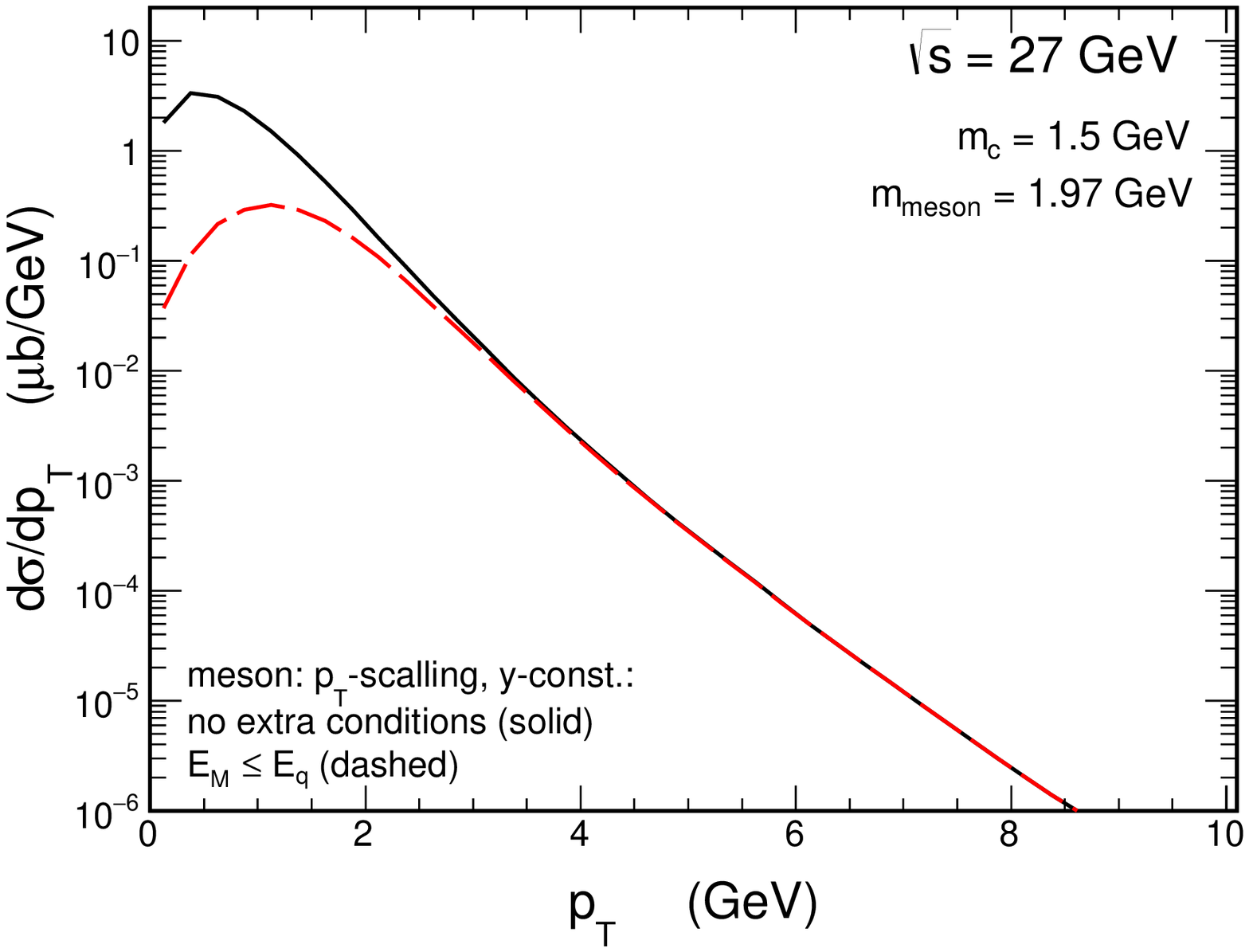}}
\end{minipage}
\caption{
\small
Transverse momentum distribution of $D_s^{\pm}$ mesons calculated in the standard approach
with (dashed) and without (solid) the $E_{D} \leq E_{q}$ condition for $\sqrt{s} = 13$ TeV (left) and $\sqrt{s} = 27$ GeV.
Here the Peterson fragmentation function was used.
}
\label{energy_cons}
\end{figure}

We start presentation of numerical results with the illustration of the violation of the "energy conservation" mentioned above
in the case of the standard approach for massive-to-massive
hadronization with unchanged rapidity scenario. In Fig.~\ref{energy_cons} we observe that the results of the standard calculations
with (dashed lines) and without (solid lines) the $E_{D} \leq E_{q}$ condition differ significantly. There is a huge damping of the $D$-meson
distributions at $p_{T} < 2$ GeV due to this limitation for both considered energies (left and right panels). According to our experience in the subject, such a huge effect is definately not supported \textit{e.g.} by the LHC charm data. It clearly shows a strong limitation of the applicability of the standard approach which can be safely used only when the hadron transverse momentum is large enough (larger than hadron mass).

\begin{figure}[!h]
\begin{minipage}{0.47\textwidth}
  \centerline{\includegraphics[width=1.0\textwidth]{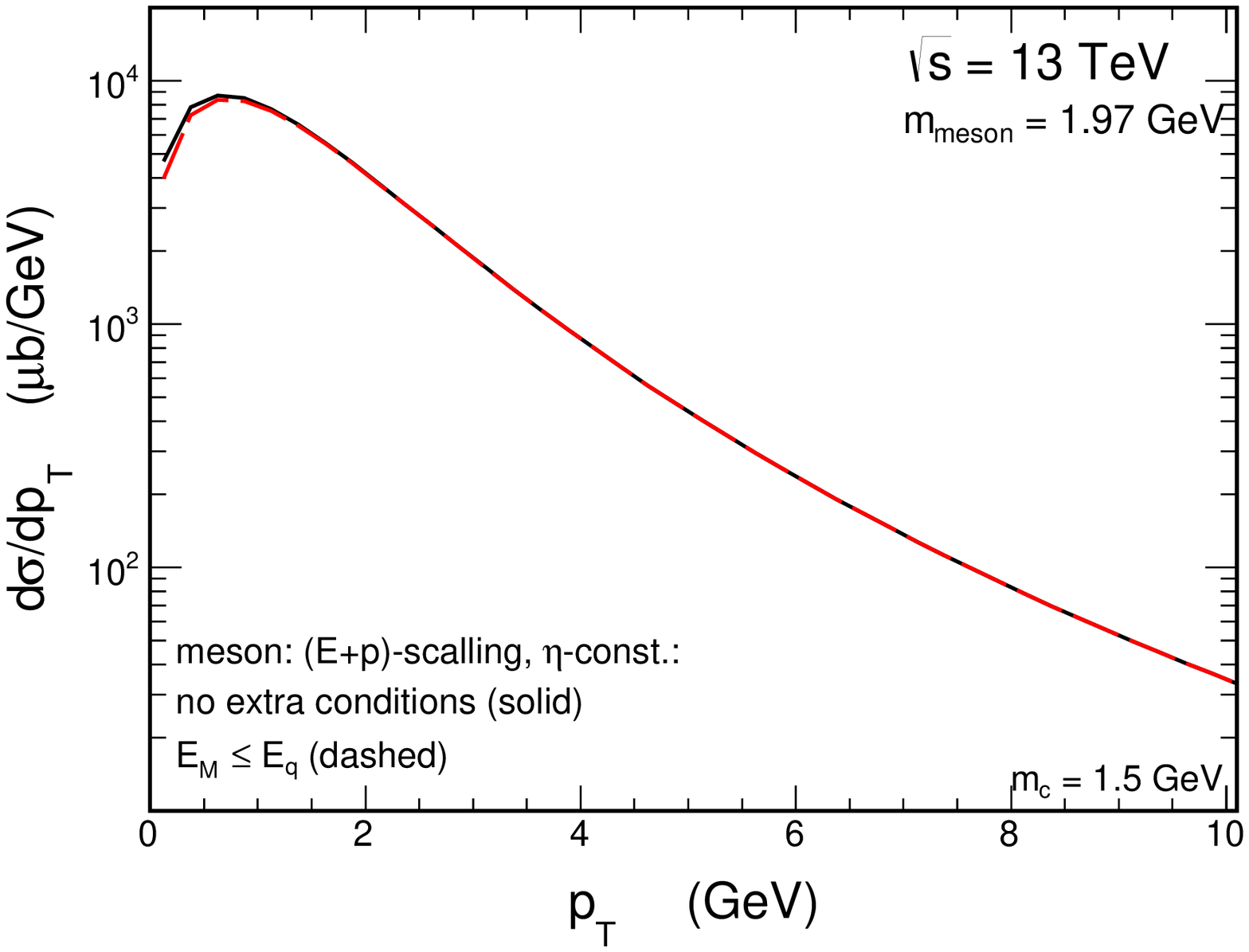}}
\end{minipage}
\begin{minipage}{0.47\textwidth}
  \centerline{\includegraphics[width=1.0\textwidth]{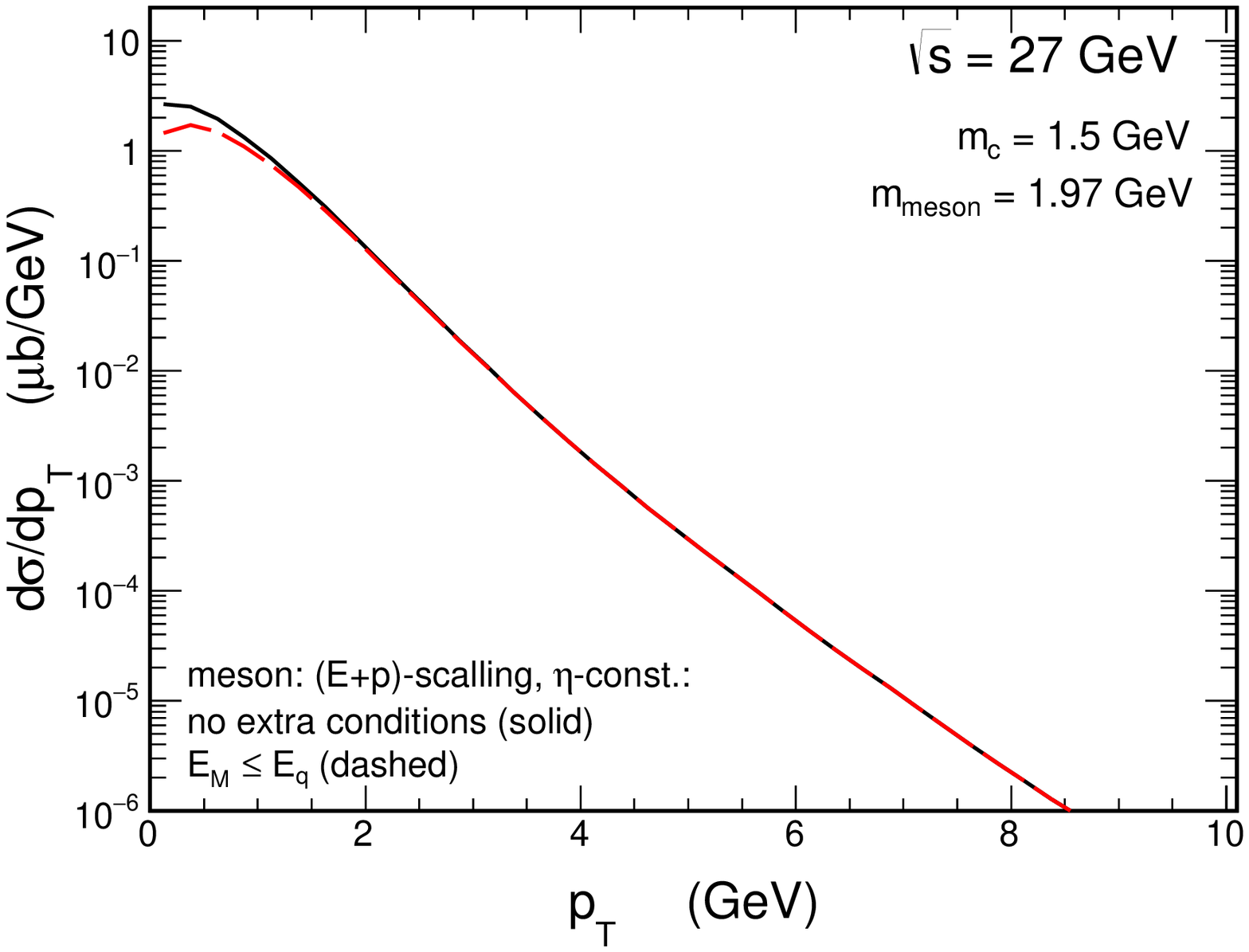}}
\end{minipage}
\caption{
\small
Transverse momentum distribution of $D_s^{\pm}$ mesons calculated with the light-cone scaling
with (dashed) and without (solid) the $E_{D} \leq E_{q}$ condition for $\sqrt{s} = 13$ TeV (left) and $\sqrt{s} = 27$ GeV.
Here the Peterson fragmentation function was used.
}
\label{energy_cons_Ppscal}
\end{figure}

The situation changes when we apply our framework with constant emission angle and with one of the proposed scaling procedure, e.g. the light-cone scaling. In Fig.~\ref{energy_cons_Ppscal} we show the corresponding results again with (dashed lines) and without (solid lines) the $E_{D} \leq E_{q}$ condition. In this case both results seem to coincide even at the low c.m.s. collision energy. It means that this prescription
seem to satisfy the "energy conservation" also in the small transverse momentum regime. Moreover, this model leads to results consistent with the standard approach calculations obtained without the extra energy condition. As a consequence, the new model predictions should not stay in contrast with LHC experimental results.        

Here and in the following, to compare shapes of quarks (and antiquarks) distributions with those
for the $D_s^{\pm}$ mesons we do not multiply the meson distributions
by the relevant fragmentation probabilities. To make the results properly normalized one needs
to multiply the presented distributions for mesons
by fragmentation probabilities $\mathrm{P}(c/\bar{c} \to D_{s}^{\pm}) = 0.56$ and 
$\mathrm{P}(s/\bar{s} \to D_{s}^{\mp}) = 0.07$ (see Ref.~\cite{Goncalves:2018zzf}).

\begin{figure}[!h]
\begin{minipage}{0.47\textwidth}
  \centerline{\includegraphics[width=1.0\textwidth]{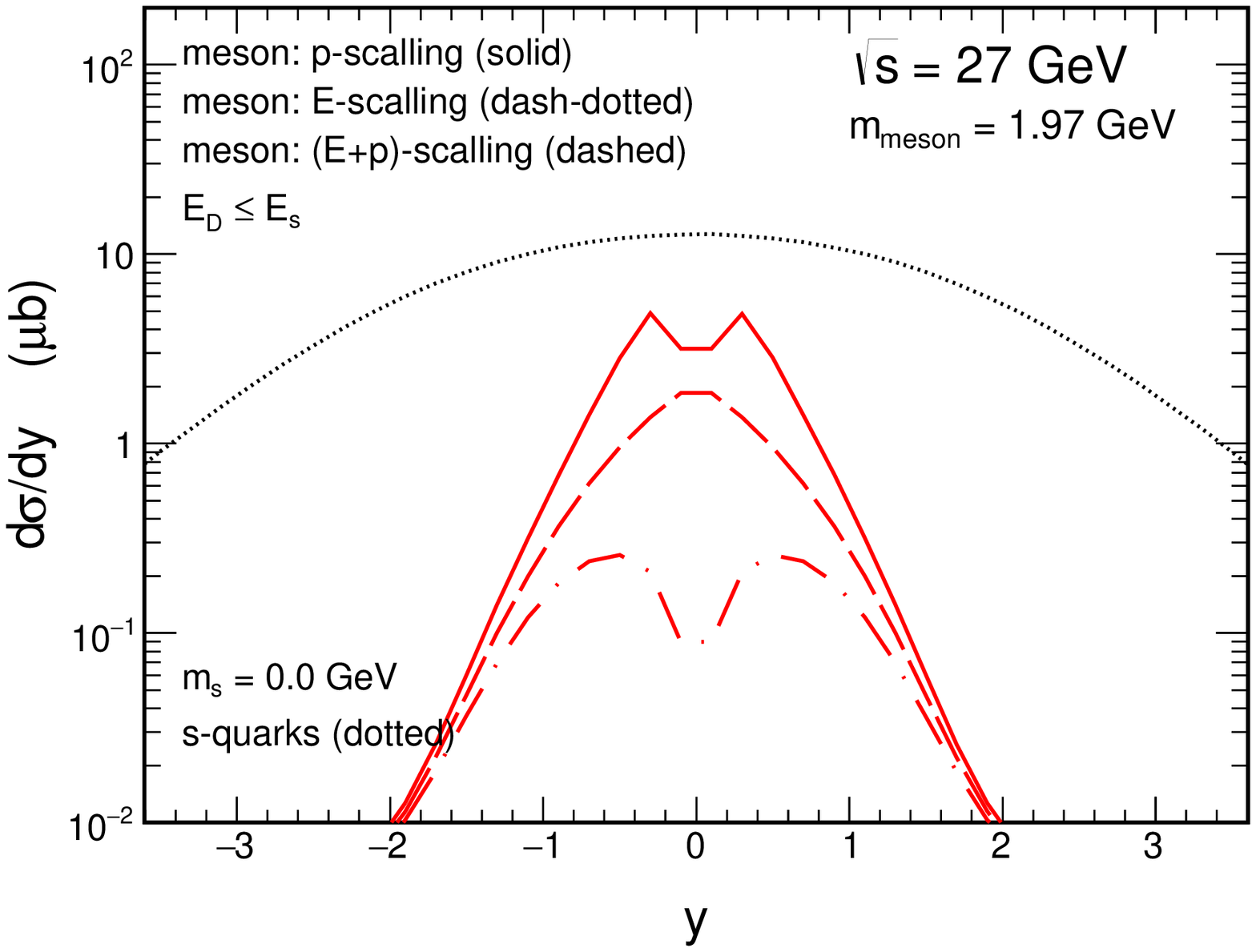}}
\end{minipage}
\begin{minipage}{0.47\textwidth}
  \centerline{\includegraphics[width=1.0\textwidth]{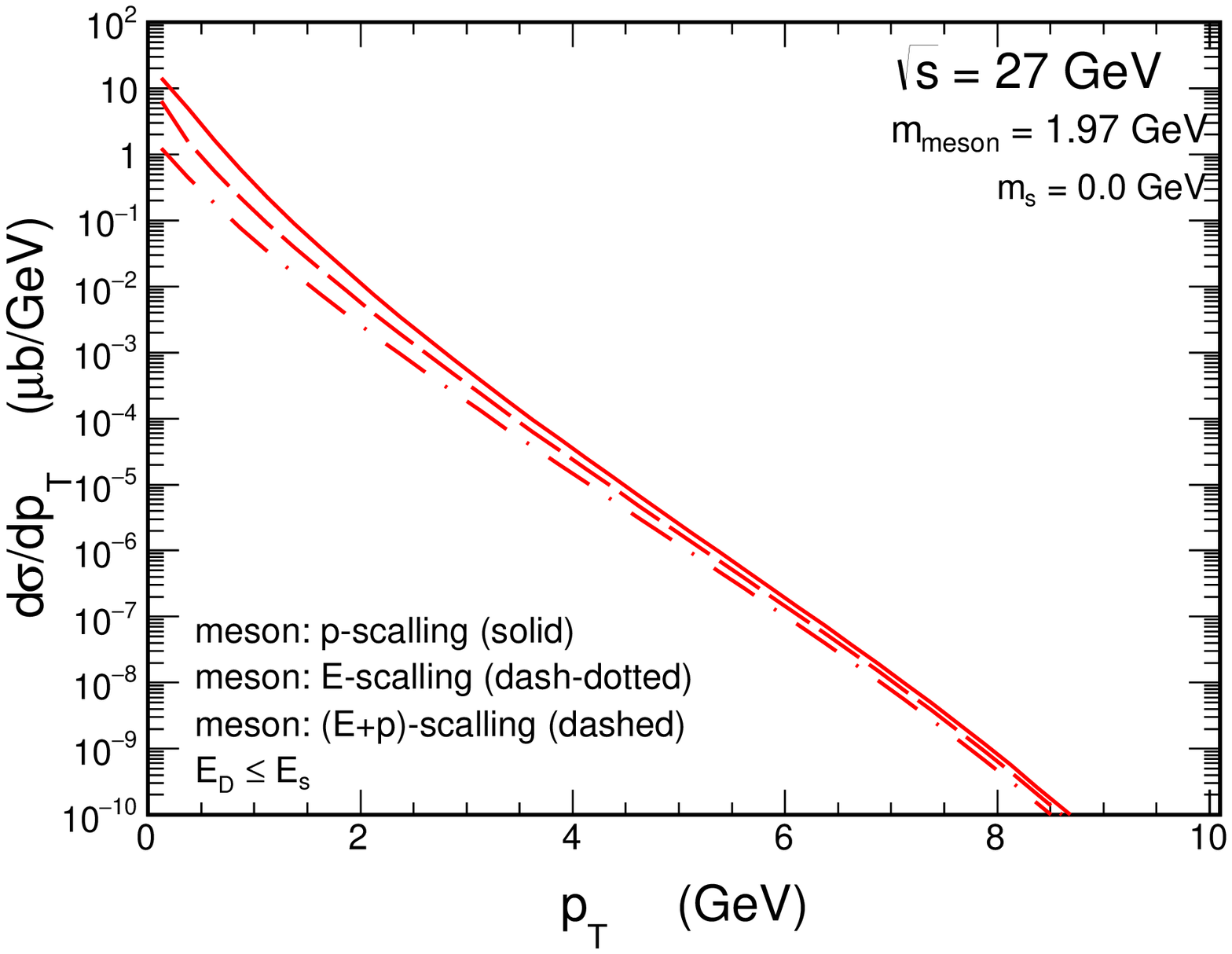}}
\end{minipage}\\
\begin{minipage}{0.47\textwidth}
  \centerline{\includegraphics[width=1.0\textwidth]{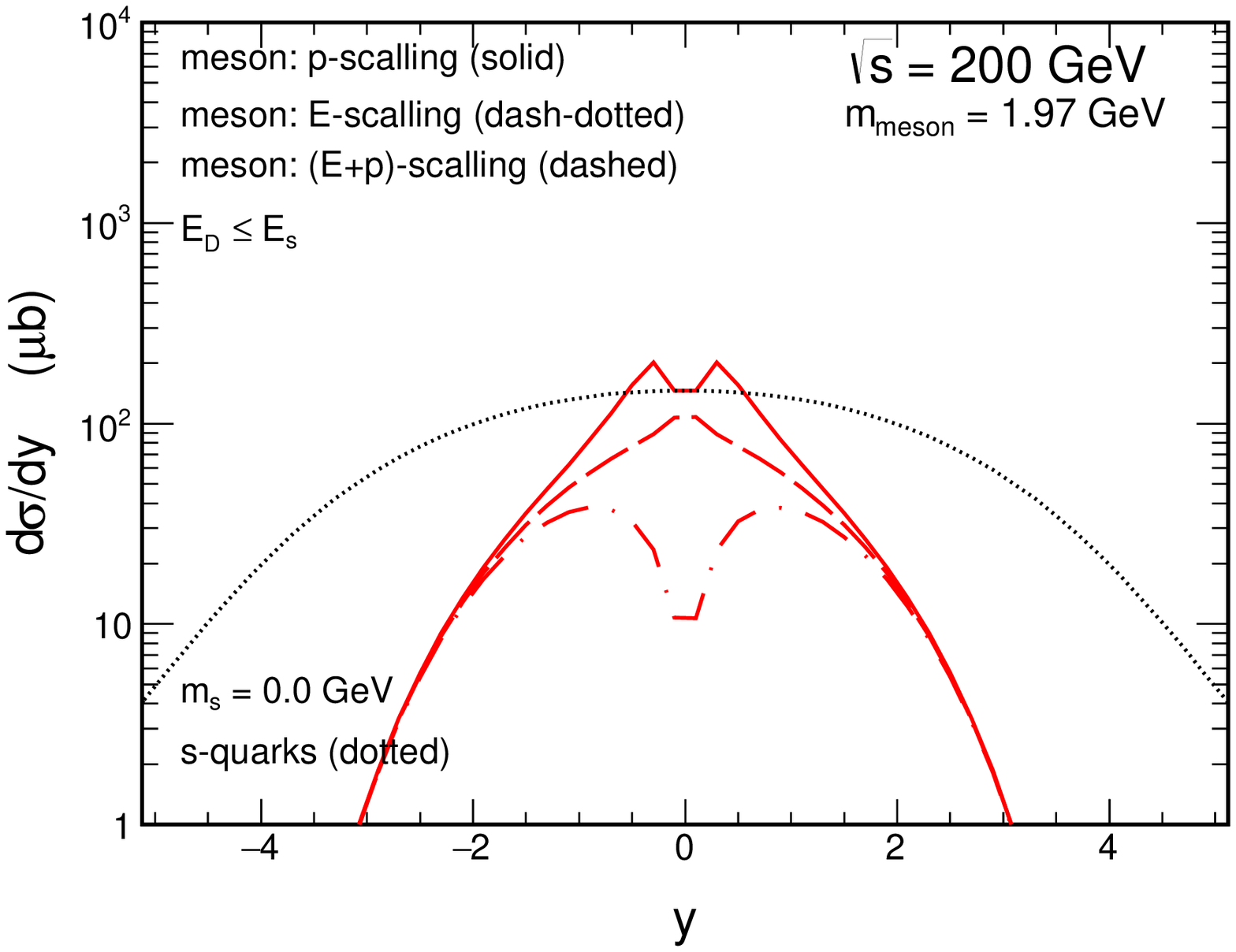}}
\end{minipage}
\begin{minipage}{0.47\textwidth}
  \centerline{\includegraphics[width=1.0\textwidth]{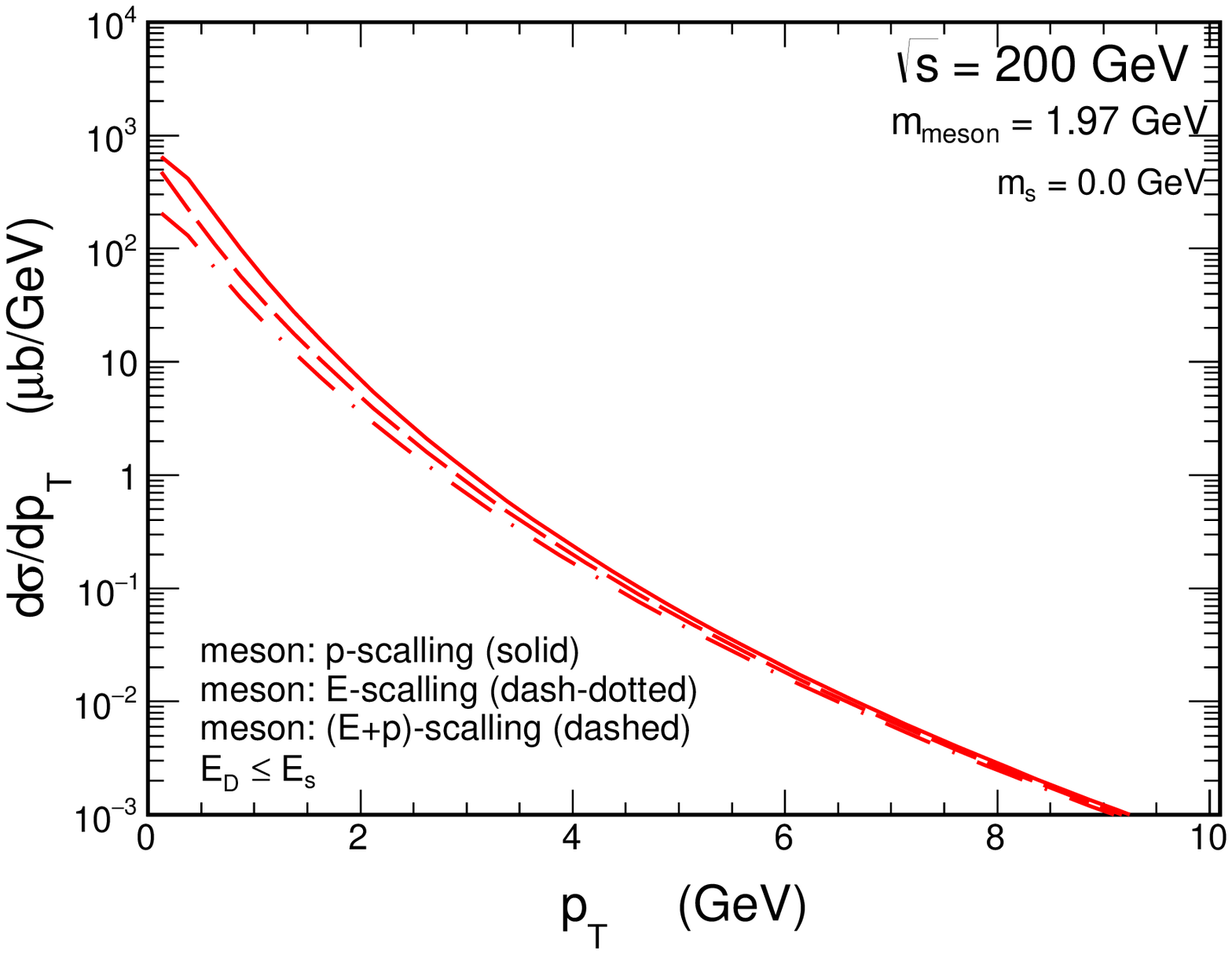}}
\end{minipage}\\
\begin{minipage}{0.47\textwidth}
  \centerline{\includegraphics[width=1.0\textwidth]{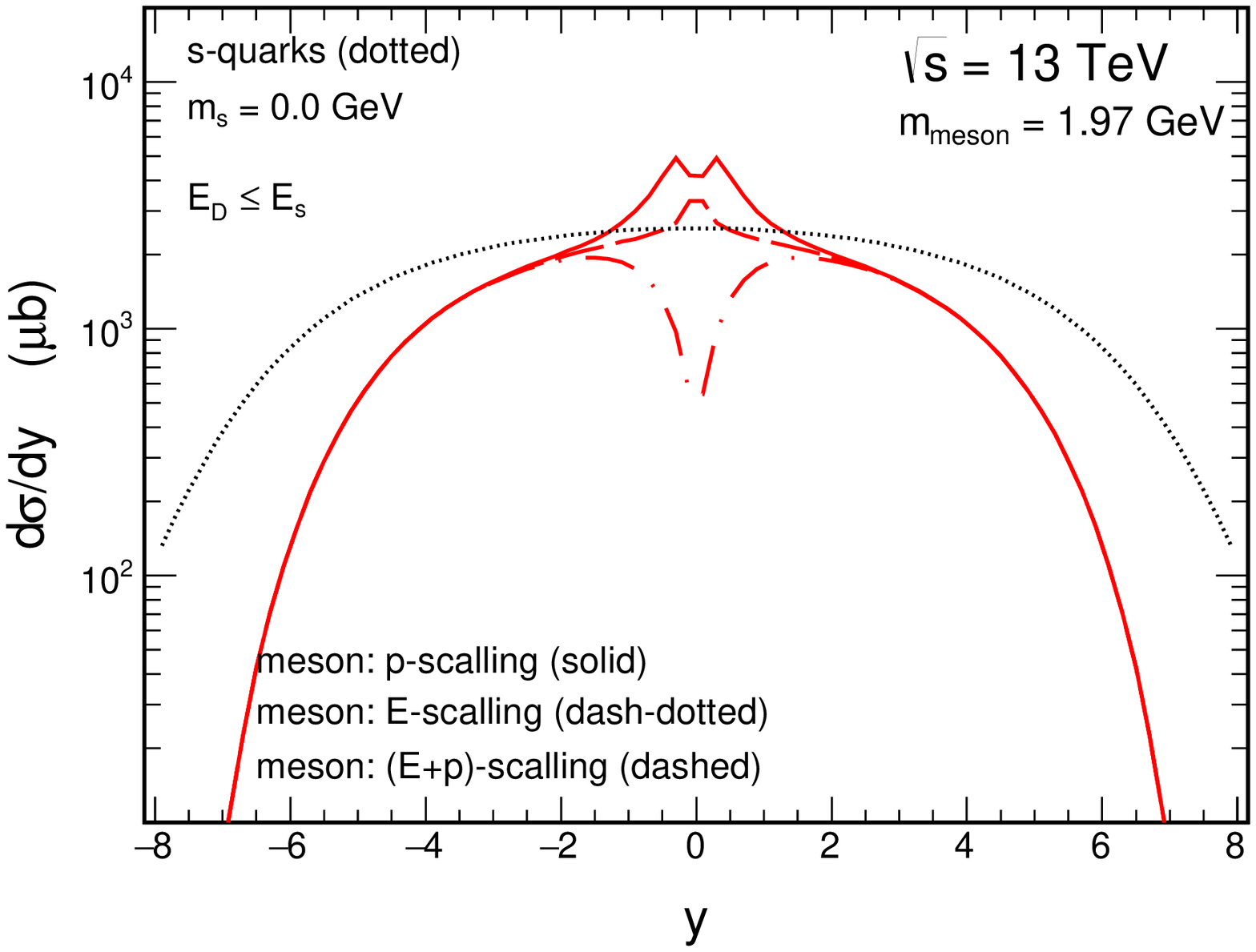}}
\end{minipage}
\begin{minipage}{0.47\textwidth}
  \centerline{\includegraphics[width=1.0\textwidth]{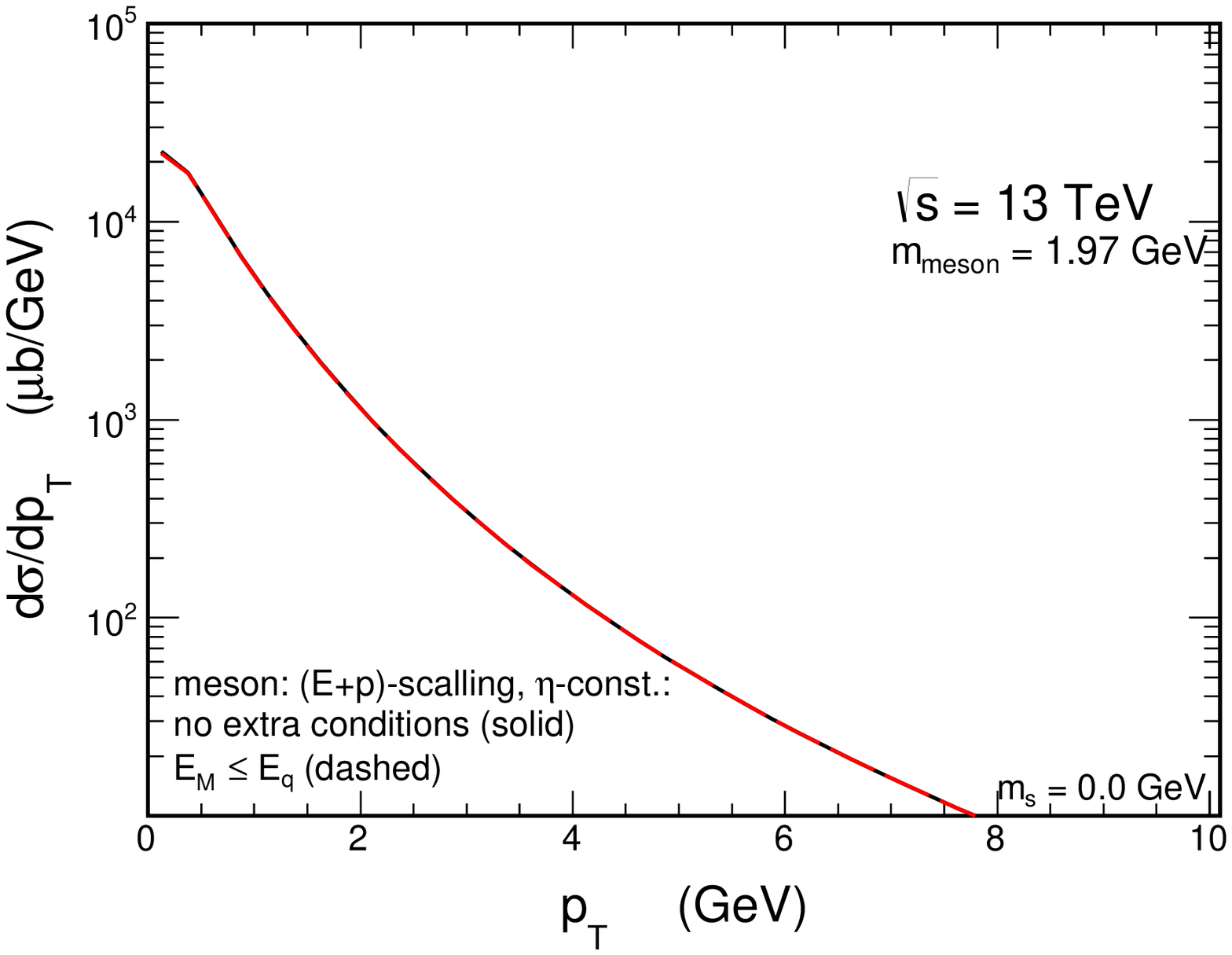}}
\end{minipage}
\caption{
\small
Rapidity (left) and transverse momentum (right) distribution 
of $D_s^{\pm}$ mesons from $s \to D_s^- / \bar s \to D_s^+$ fragmentation 
for different scaling procedures described above.
Here the reversed Peterson fragmentation function and $P(s/\bar{s} \to D^{\mp}_{s})$ = 1 were used.
}
\label{dsig_dy_light_to_Ds}
\end{figure}

In Fig.~\ref{dsig_dy_light_to_Ds} we show our results
for light-to-heavy fragmentation $s /\bar s \to D_s^{\mp}$.
The left and right panels present rapidity and transverse momentum distributions, respectively,
for $D_{s}$-meson calculated with momentum (solid lines), energy (dash-doted lines) and light-cone
(dashed lines) scaling method. The top, middle and bottom panels show results for different energies: $\sqrt{s} = 27$ GeV, $\sqrt{s} = 200$ GeV and $\sqrt{s} = 13$ TeV, respectively. The dotted lines correspond to the standard approach with transverse momentum scaling and with unchanged (parton $\to$ hadron) rapidity. 
The different approaches lead to quite different results. The new approaches differ significantly from the standard one. The discrepancy increases when going to low energy regime. We observe expected shift of the cross section for meson from the forward/backward regions to midrapidity with respect to the quark distribution. However, within the new approaches some problematic behaviour appears at midrapidites of meson. It is strictly related to the region of small meson transverse momenta where mass effects, even at high energies, play non-negligible role. The transverse momentum distributions for momentum scaling coincide with the distributions calculated with the standard approach when considering the whole rapidity range.

\begin{figure}[!h]
\begin{minipage}{0.47\textwidth}
  \centerline{\includegraphics[width=1.0\textwidth]{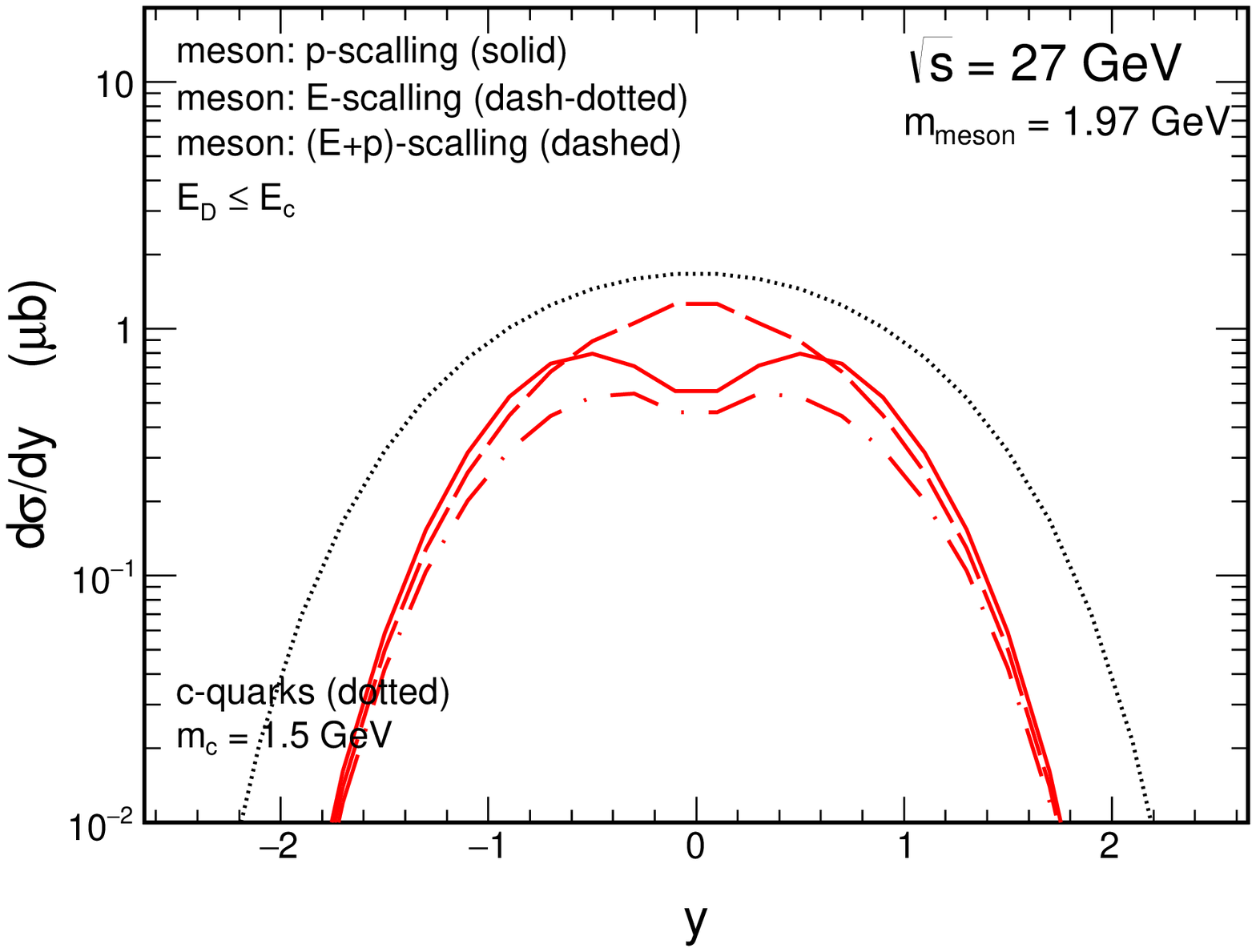}}
\end{minipage}
\begin{minipage}{0.47\textwidth}
  \centerline{\includegraphics[width=1.0\textwidth]{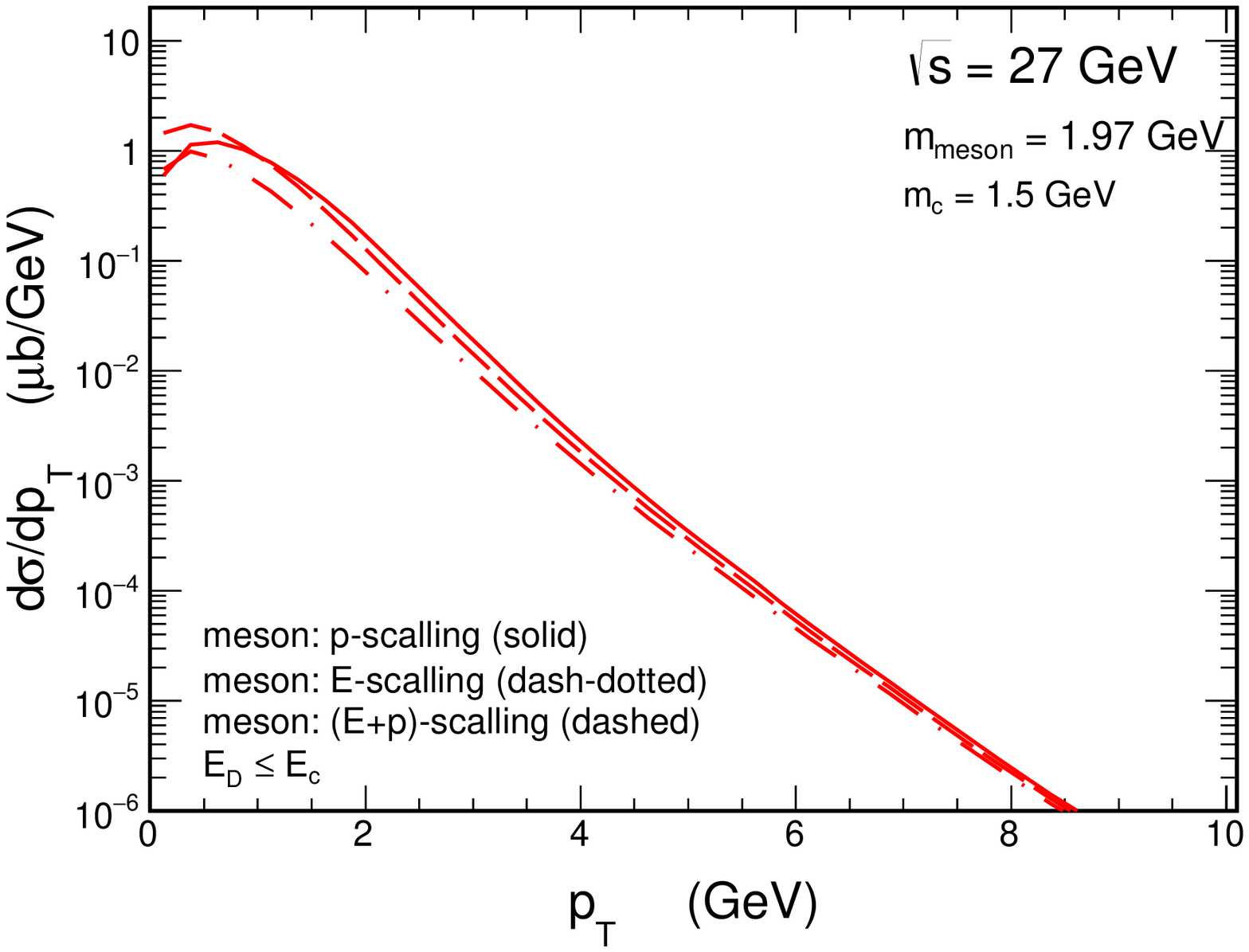}}
\end{minipage}\\
\begin{minipage}{0.47\textwidth}
  \centerline{\includegraphics[width=1.0\textwidth]{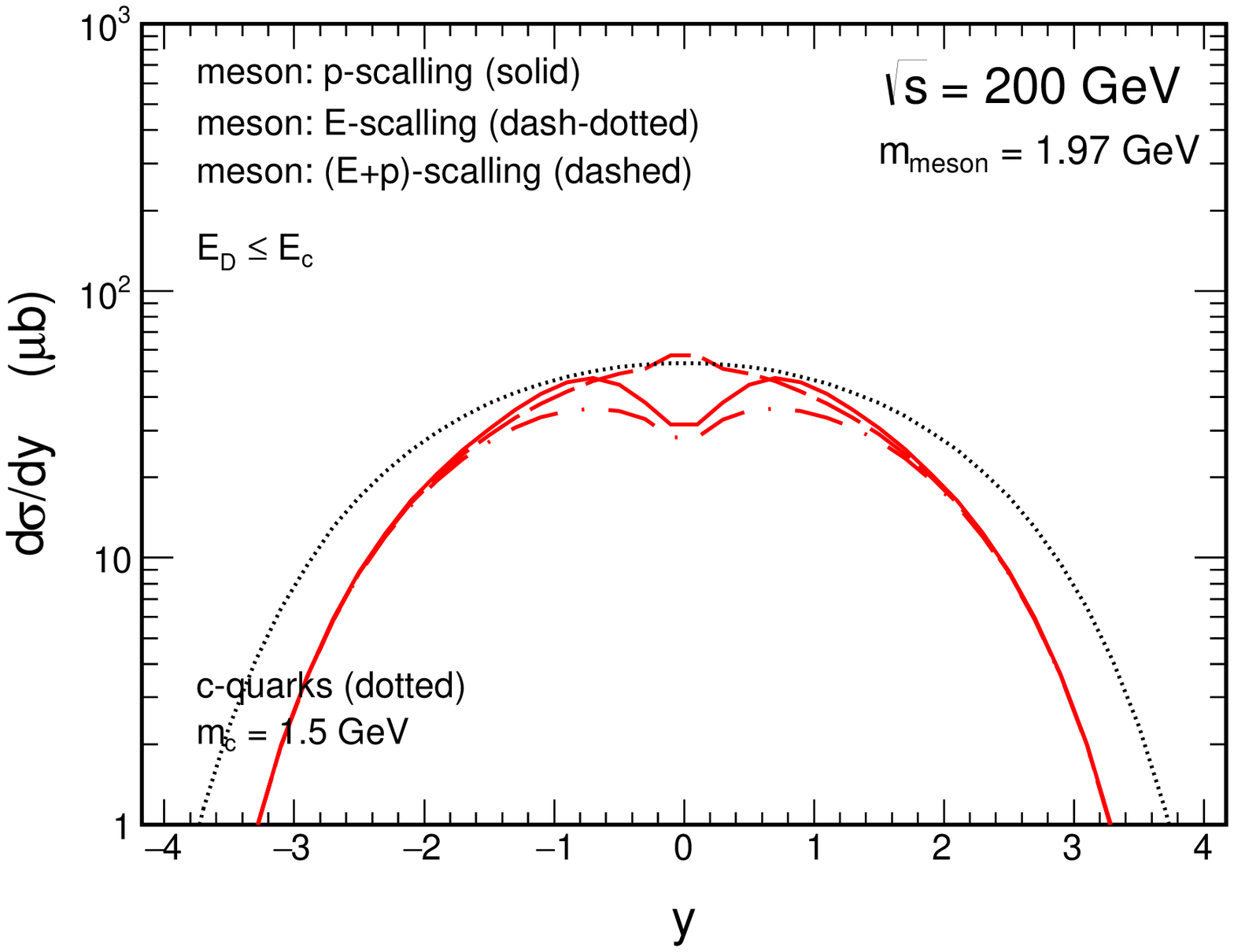}}
\end{minipage}
\begin{minipage}{0.47\textwidth}
  \centerline{\includegraphics[width=1.0\textwidth]{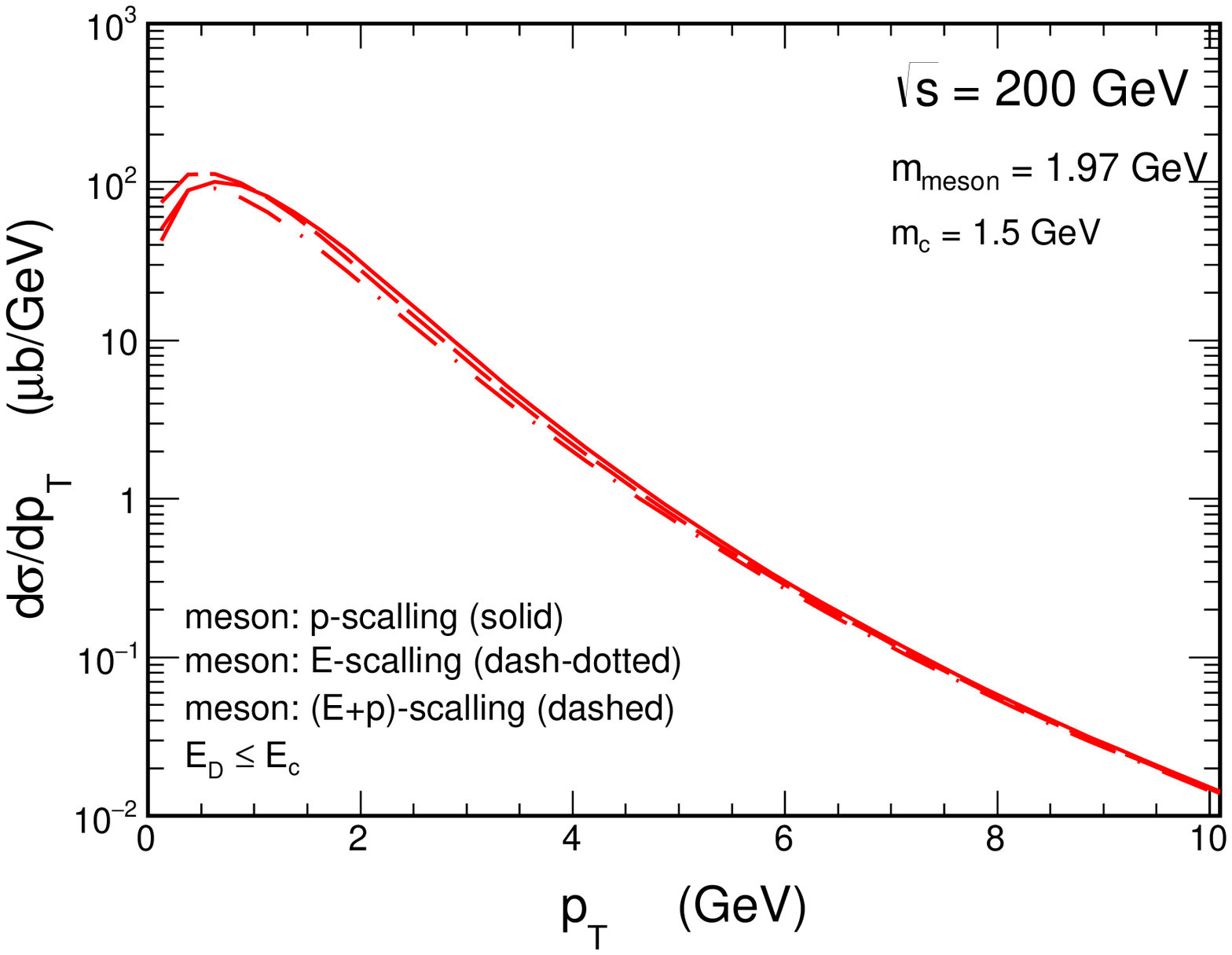}}
\end{minipage}\\
\begin{minipage}{0.47\textwidth}
  \centerline{\includegraphics[width=1.0\textwidth]{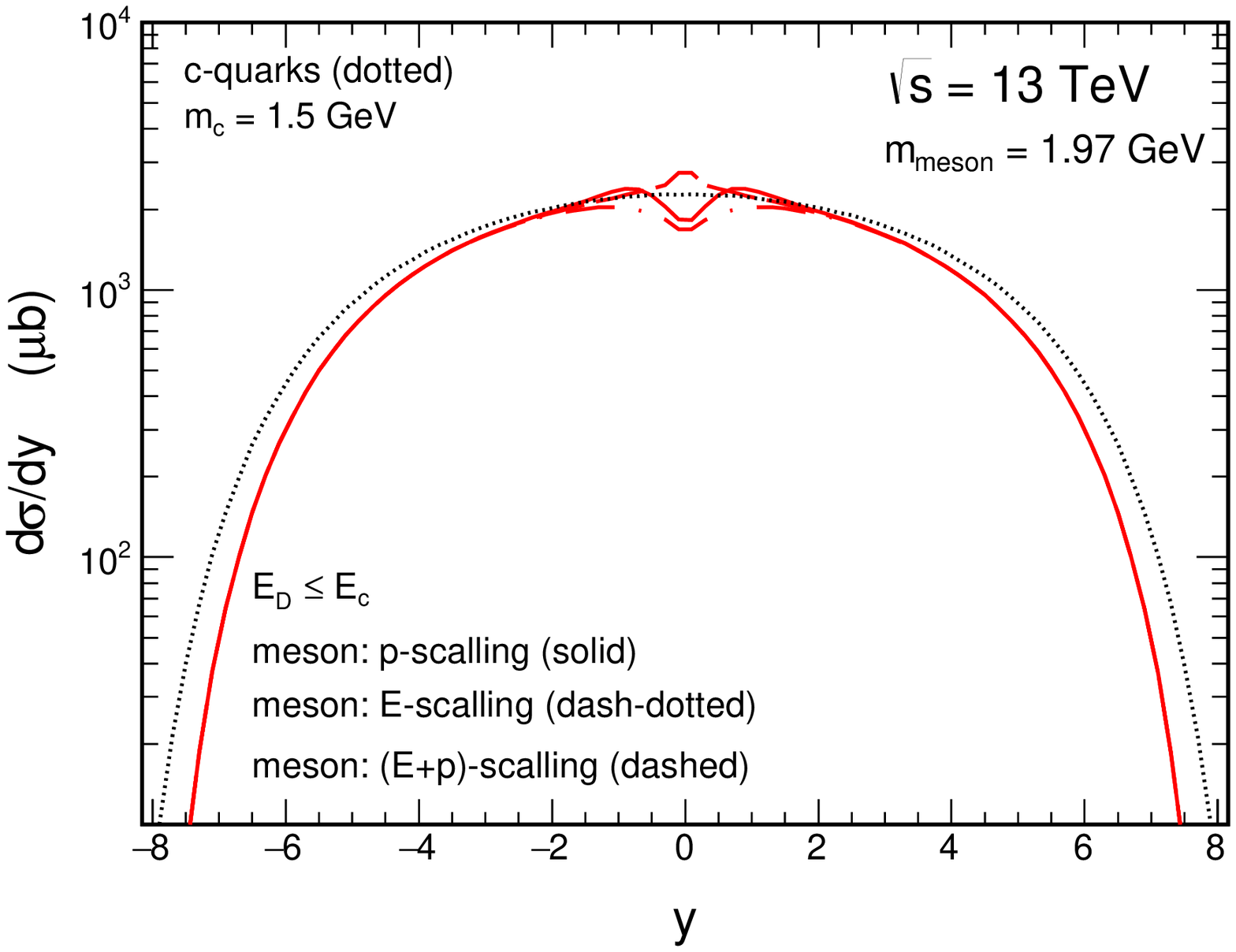}}
\end{minipage}
\begin{minipage}{0.47\textwidth}
  \centerline{\includegraphics[width=1.0\textwidth]{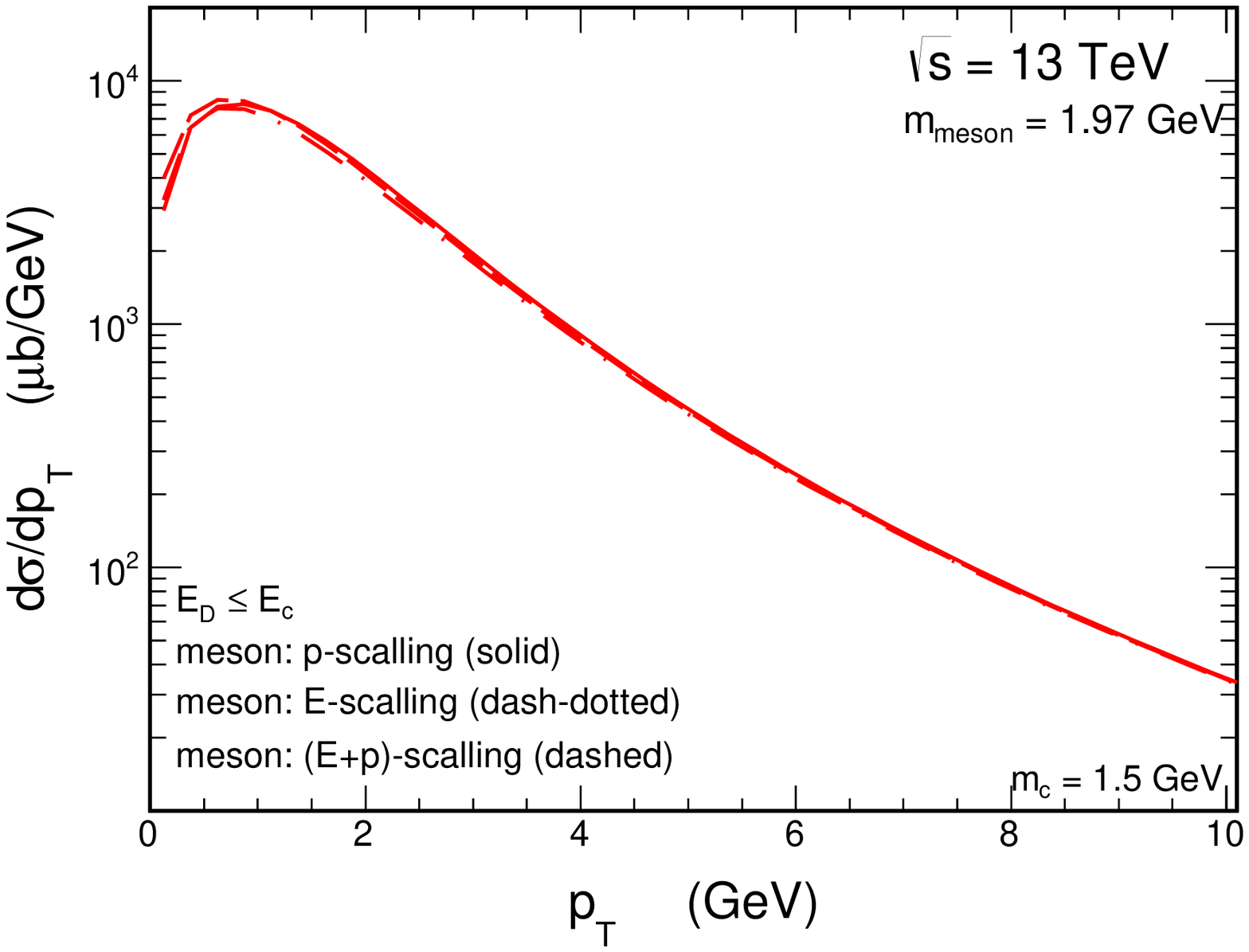}}
\end{minipage}
\caption{\small
Rapidity (left) and transverse momentum (right) distributions 
of $D_s^{\pm}$ mesons from $c/\bar c \to D_s$ fragmentation for different scalings 
described above.
Here the Peterson fragmentation function is used with $P(c/\bar{c} \to D^{\pm}_{s})$ = 1.
}
\label{dsig_dy_heavy_to_Ds}
\end{figure}

In Fig.~\ref{dsig_dy_heavy_to_Ds} we show our results
for heavy-to-heavy fragmentation $c /\bar c \to D_s^{\pm}$.
Again, the distributions for $D_s^{\pm}$ mesons were not multiplied here
by the $P(c/\bar{c} \to D^{\pm}_{s})$ probability to concentrate on 
their shapes only.
The differences between the different approaches here are 
much smaller than for light-to-heavy fragmentation which is 
caused by the fact that masses of $c/\bar c$
are similar to masses of $D_s^{\pm}$ mesons.

\begin{figure}[!h]
\begin{minipage}{0.47\textwidth}
  \centerline{\includegraphics[width=1.0\textwidth]{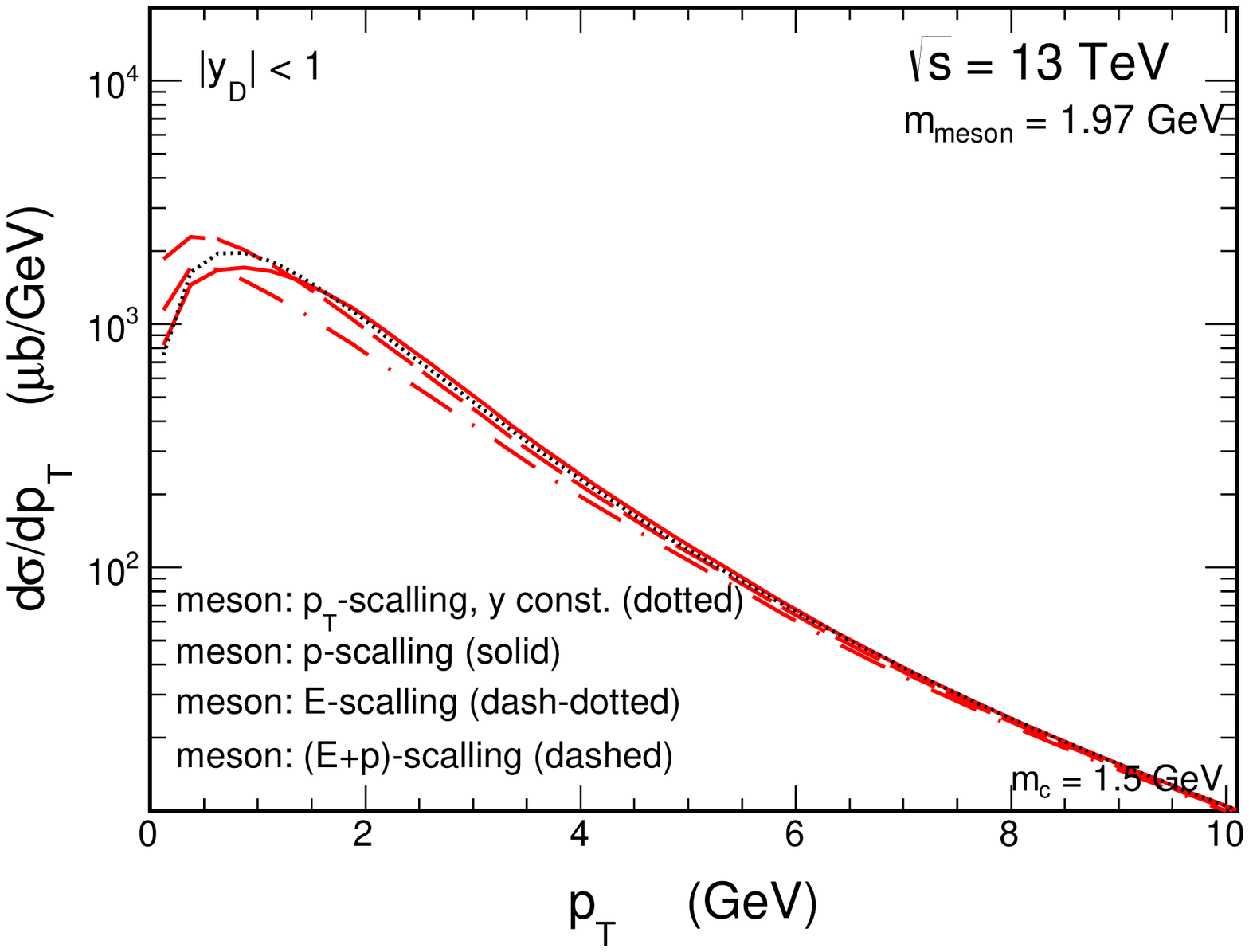}}
\end{minipage}
\begin{minipage}{0.47\textwidth}
  \centerline{\includegraphics[width=1.0\textwidth]{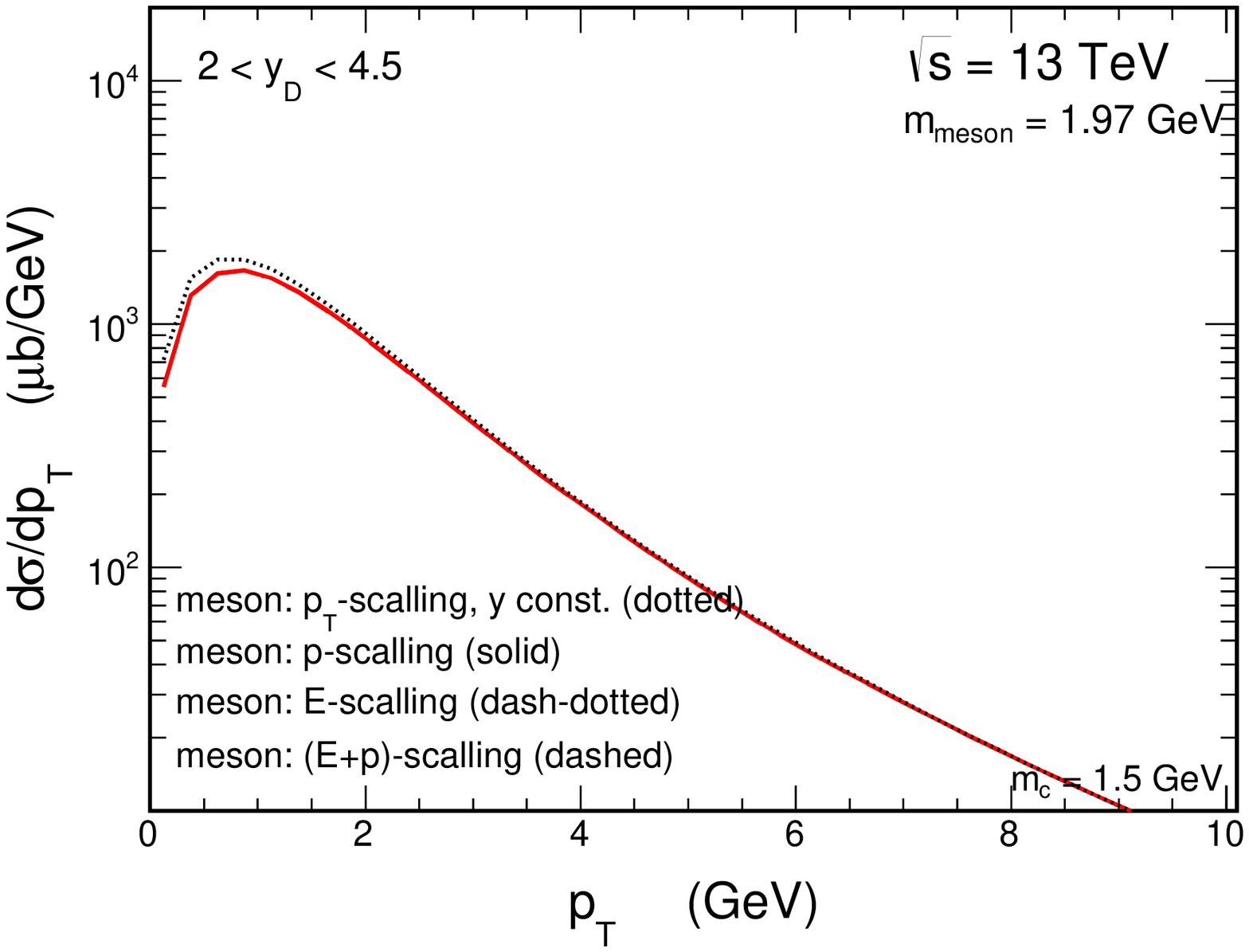}}
\end{minipage}
\caption{\small
Transverse momentum distributions of $D_s^{\pm}$ mesons from 
$c/{\bar c} \to D_s^{\pm}$ fragmentation for $\sqrt{s} = 13$ TeV for two regions of rapidities:
relevant for ALICE (left) and relevant for LHCb (right).
}
\label{fig:dsig_dpt_heavy_to_Ds_limitedy}
\end{figure}

\begin{figure}[!h]
\begin{minipage}{0.47\textwidth}
  \centerline{\includegraphics[width=1.0\textwidth]{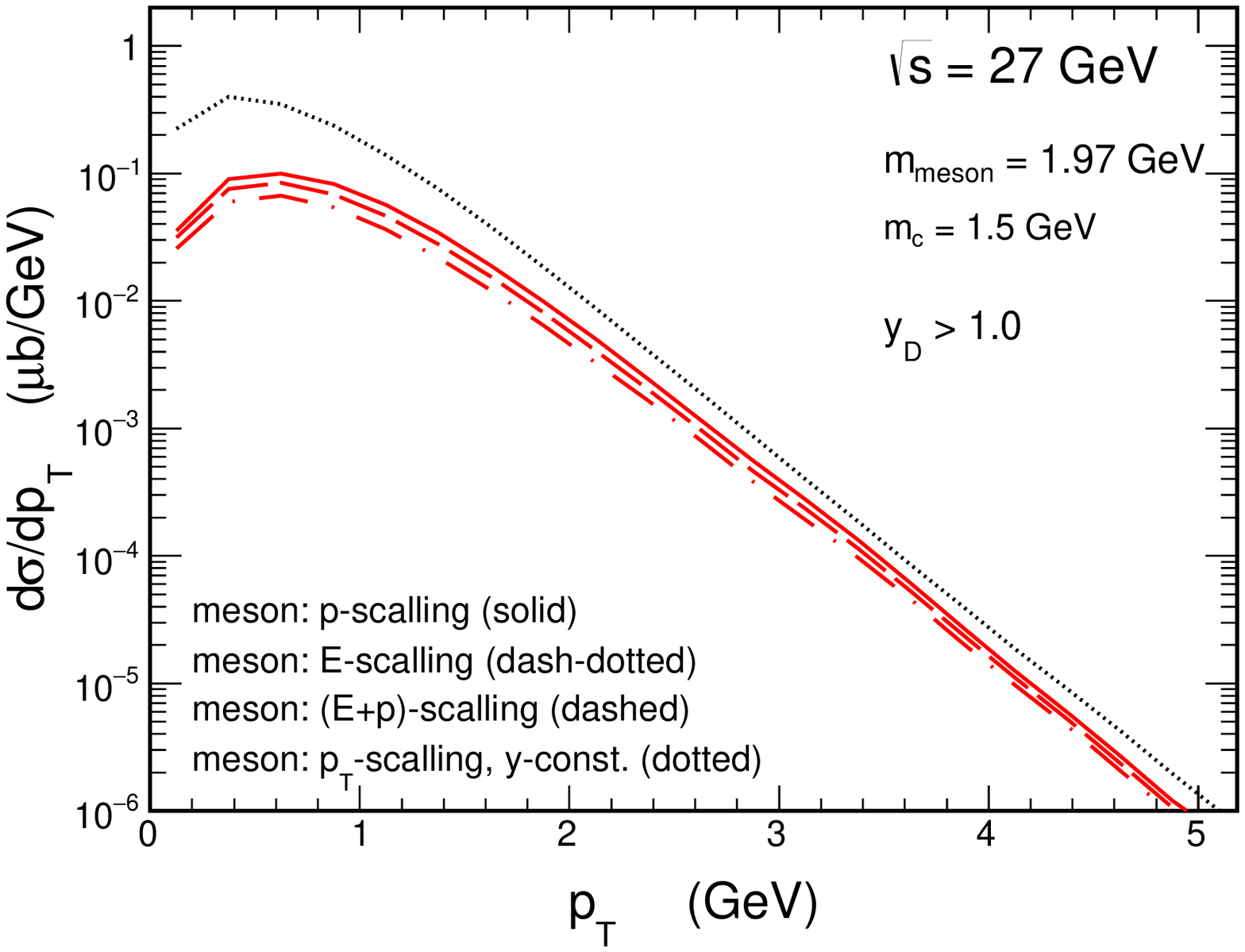}}
\end{minipage}
\begin{minipage}{0.47\textwidth}
  \centerline{\includegraphics[width=1.0\textwidth]{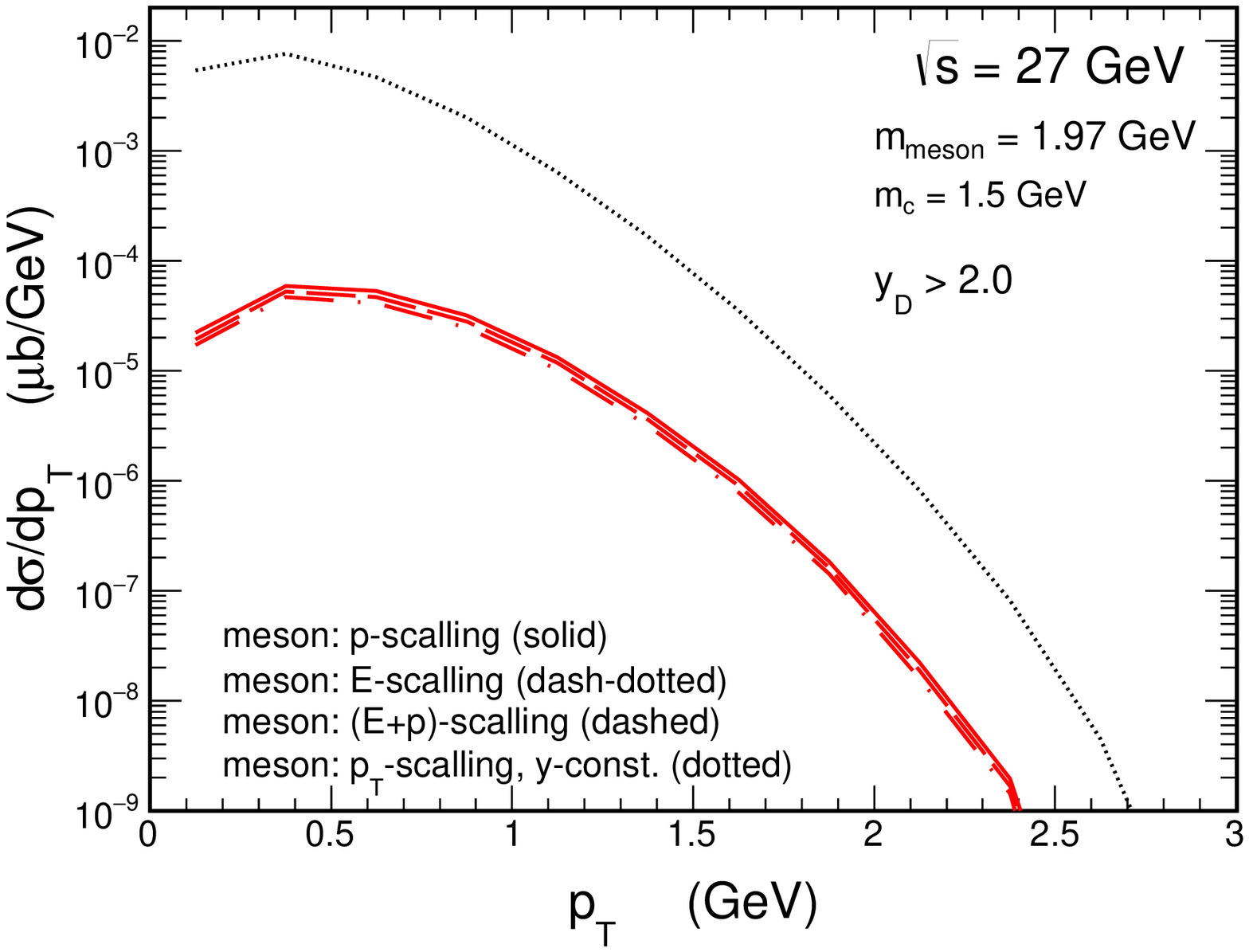}}
\end{minipage}
\caption{\small
Transverse momentum distributions of $D_s^{\pm}$ mesons from 
$c/{\bar c} \to D_s^{\pm}$ fragmentation for $\sqrt{s} = 27$ GeV for two regions of rapidities:
$y_{D} > 1$ (left) and $y_{D} > 2$ (right) .
}
\label{fig:dsig_dpt_heavy_to_Ds_limitedy_27GeV}
\end{figure}

\begin{figure}[!h]
\begin{minipage}{0.47\textwidth}
  \centerline{\includegraphics[width=1.0\textwidth]{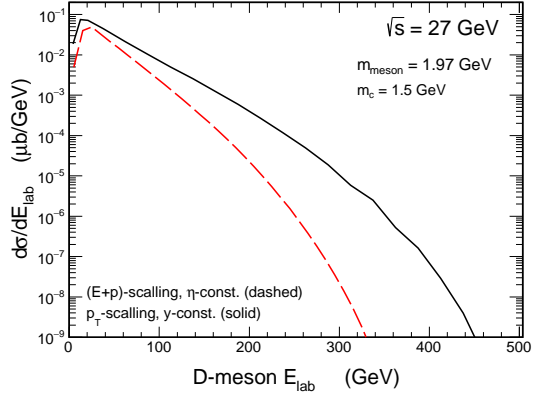}}
\end{minipage}
\caption{\small
Energy distributions of $D_s^{\pm}$ mesons in the laboratory frame from 
$c/{\bar c} \to D_s^{\pm}$ fragmentation for $\sqrt{s} = 27$ GeV.
}
\label{fig:dsig_dElab}
\end{figure}

\begin{figure}[!h]
\begin{minipage}{0.47\textwidth}
  \centerline{\includegraphics[width=1.0\textwidth]{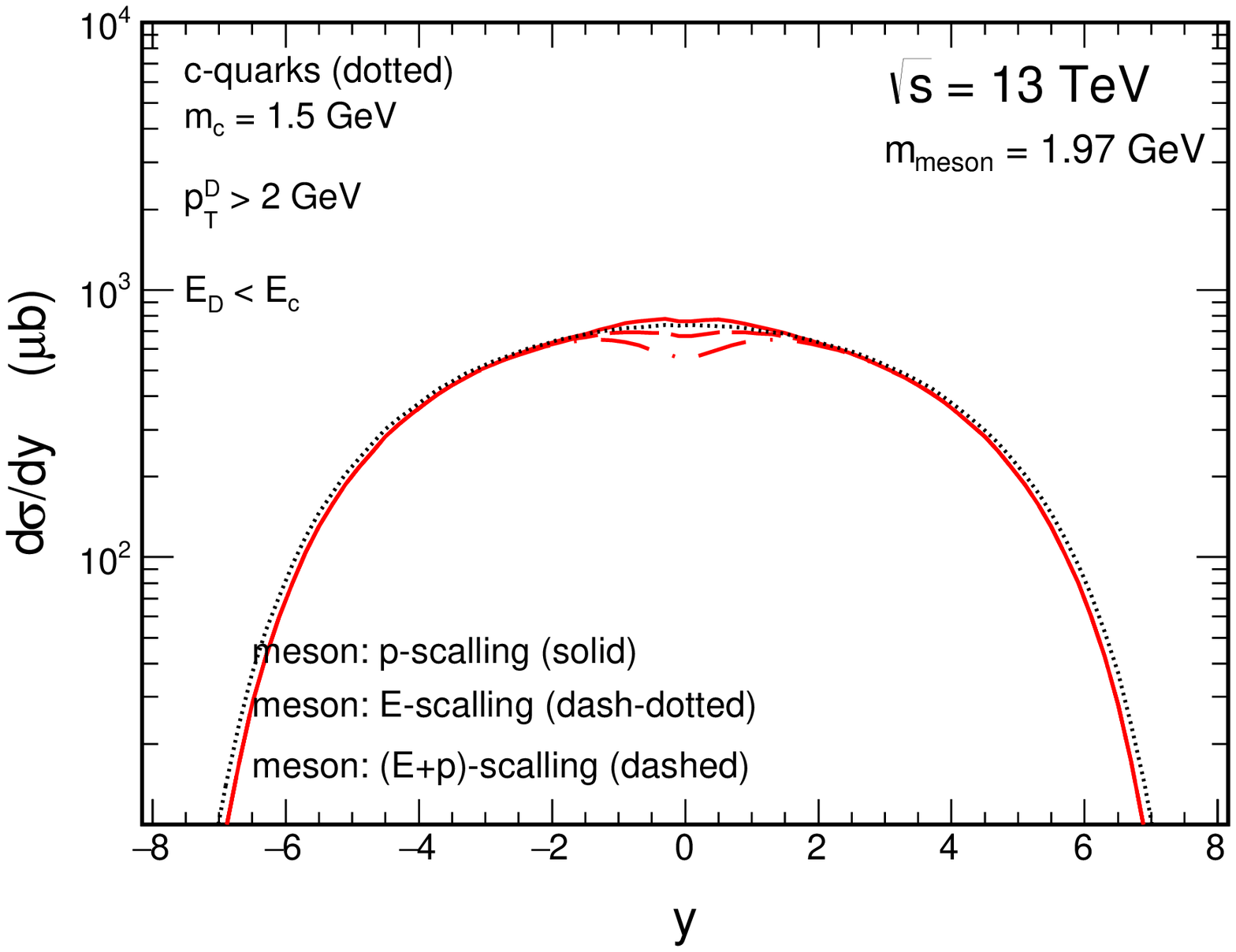}}
\end{minipage}
\begin{minipage}{0.47\textwidth}
  \centerline{\includegraphics[width=1.0\textwidth]{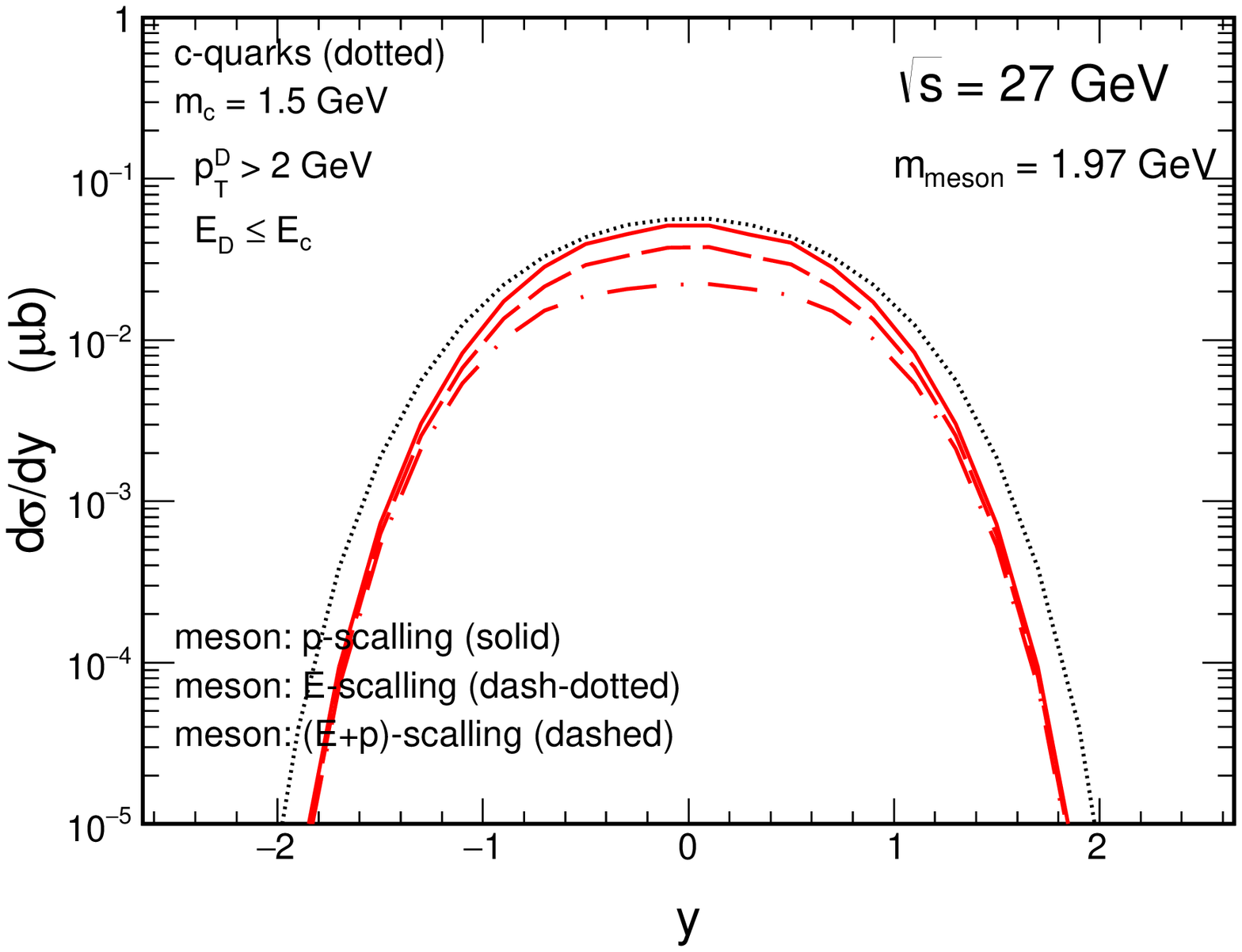}}
\end{minipage}\\
\begin{minipage}{0.47\textwidth}
  \centerline{\includegraphics[width=1.0\textwidth]{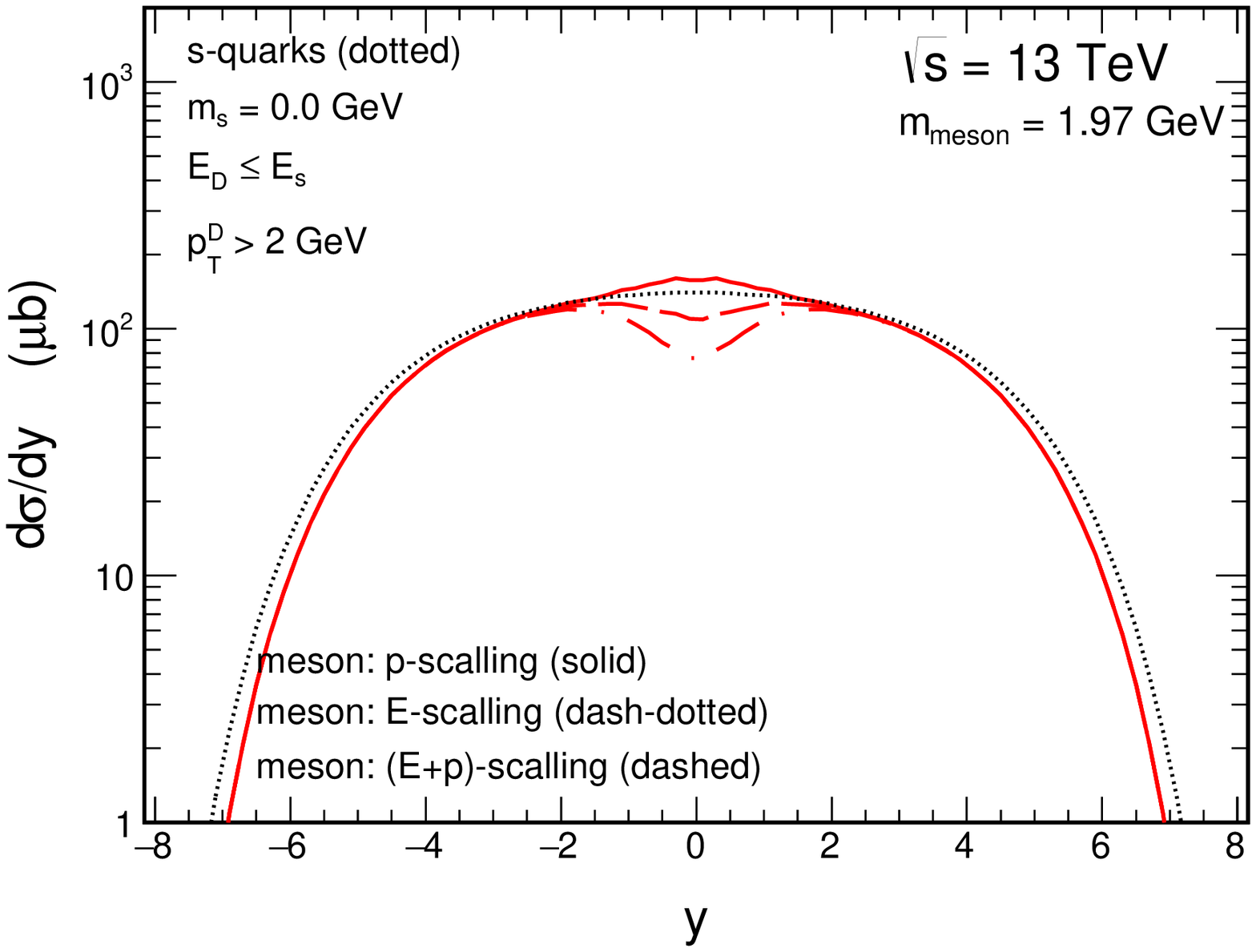}}
\end{minipage}
\begin{minipage}{0.47\textwidth}
  \centerline{\includegraphics[width=1.0\textwidth]{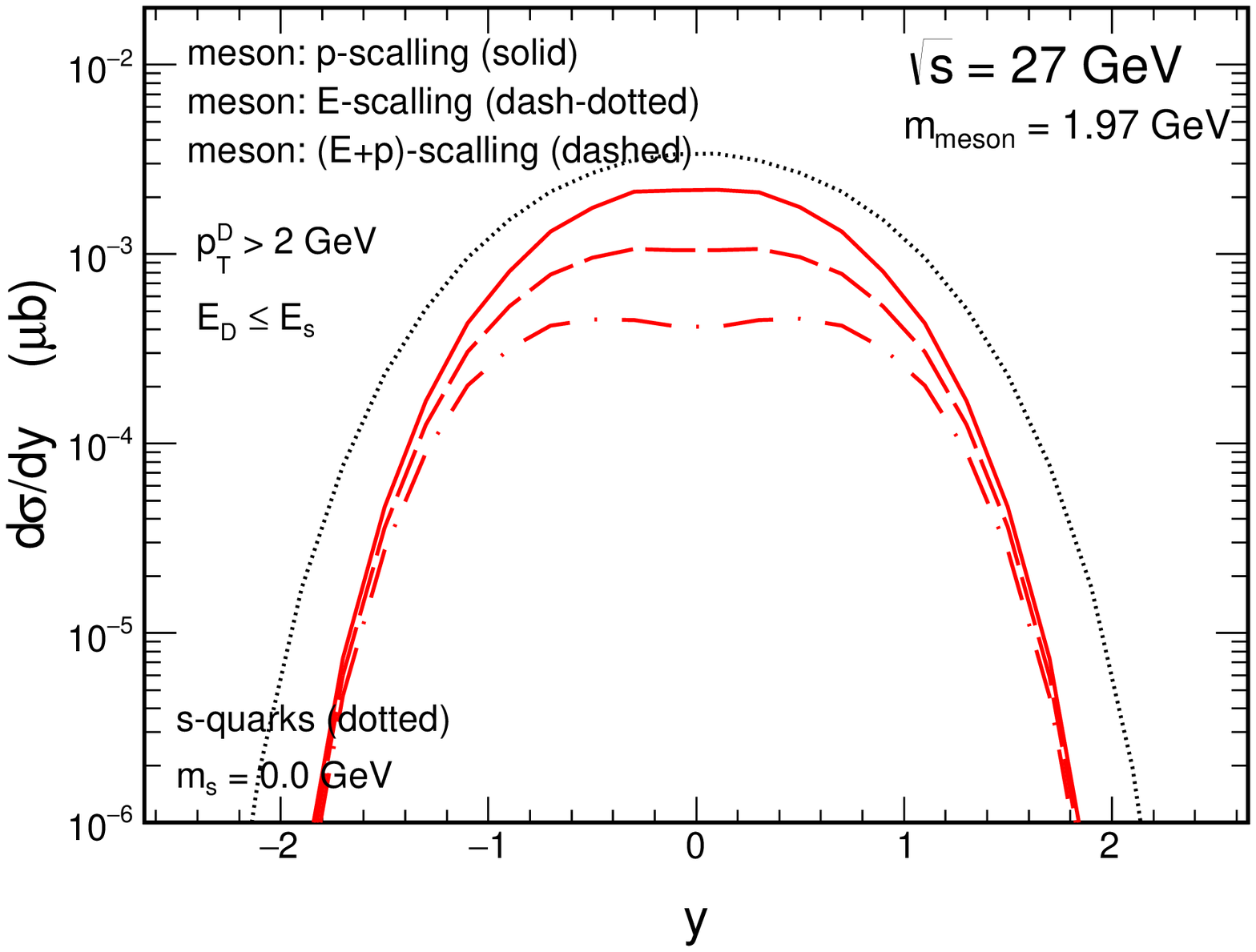}}
\end{minipage}
\caption{\small
Rapidity distributions of $D_s^{\pm}$ mesons for $c/\bar c \to
D_s^{\pm}$ (top panels) and $s/\bar s \to D_s^{\mp}$ (bottom panels)
for $\sqrt{s}$ = 13 TeV (left) and $\sqrt{s}$ = 27 GeV (right).
Here $p_{t,D_s} >$ 2 GeV.
}
\label{fig:dsig_dy_ptDgt2}
\end{figure}

In Fig.~\ref{fig:dsig_dpt_heavy_to_Ds_limitedy} we show transverse
momentum distributions of $D_s$ mesons separately for different
ranges of rapidity for $\sqrt{s} = 13$ TeV. On the left hand side we show results for
midrapidities (relevant e.g. for ALICE experiment) and on the right hand
side for forward rapidities (relevant for the LHCb experiment).
While at midrapidities the different scaling methods give rather
different results, especially for small transverse momenta, the results
of different methods coincide in forward rapidity region.

Similarly, in Fig.~\ref{fig:dsig_dpt_heavy_to_Ds_limitedy_27GeV} we show again
transverse momentum distributions of $D_s$ mesons separately for different
ranges of rapidity (left and right panel) but this time for low energy $\sqrt{s} = 27$ GeV, 
that corresponds to the planned SHIP experiment. Here, the differences between the standard and the new methods
are much more significant, especially when moving to the forward rapidities. The same conclusions can be drawn
from Fig.~\ref{fig:dsig_dElab}, where the $D_s$ meson energy distribution in the laboratory frame is shown.
The energy distribution of $D_{s}$ meson determines the energy distribution of $\nu_{\tau}/\bar{\nu_{\tau}}$ neutrinos
that could be studied at SHIP experiment in the far-forward rapidity region \cite{Bai:2018xum}.

\begin{figure}[!h]
\begin{minipage}{0.47\textwidth}
  \centerline{\includegraphics[width=1.0\textwidth]{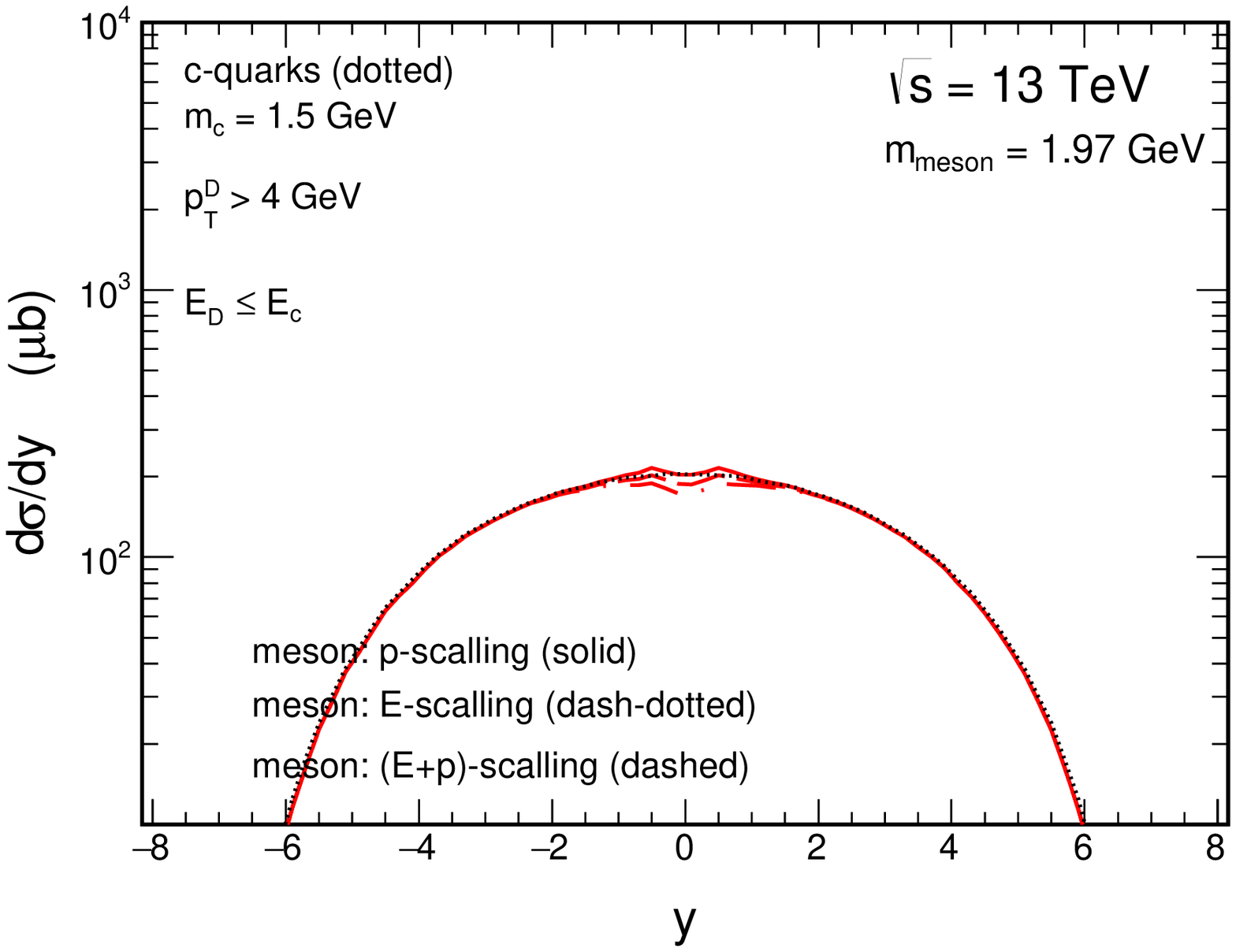}}
\end{minipage}
\begin{minipage}{0.47\textwidth}
  \centerline{\includegraphics[width=1.0\textwidth]{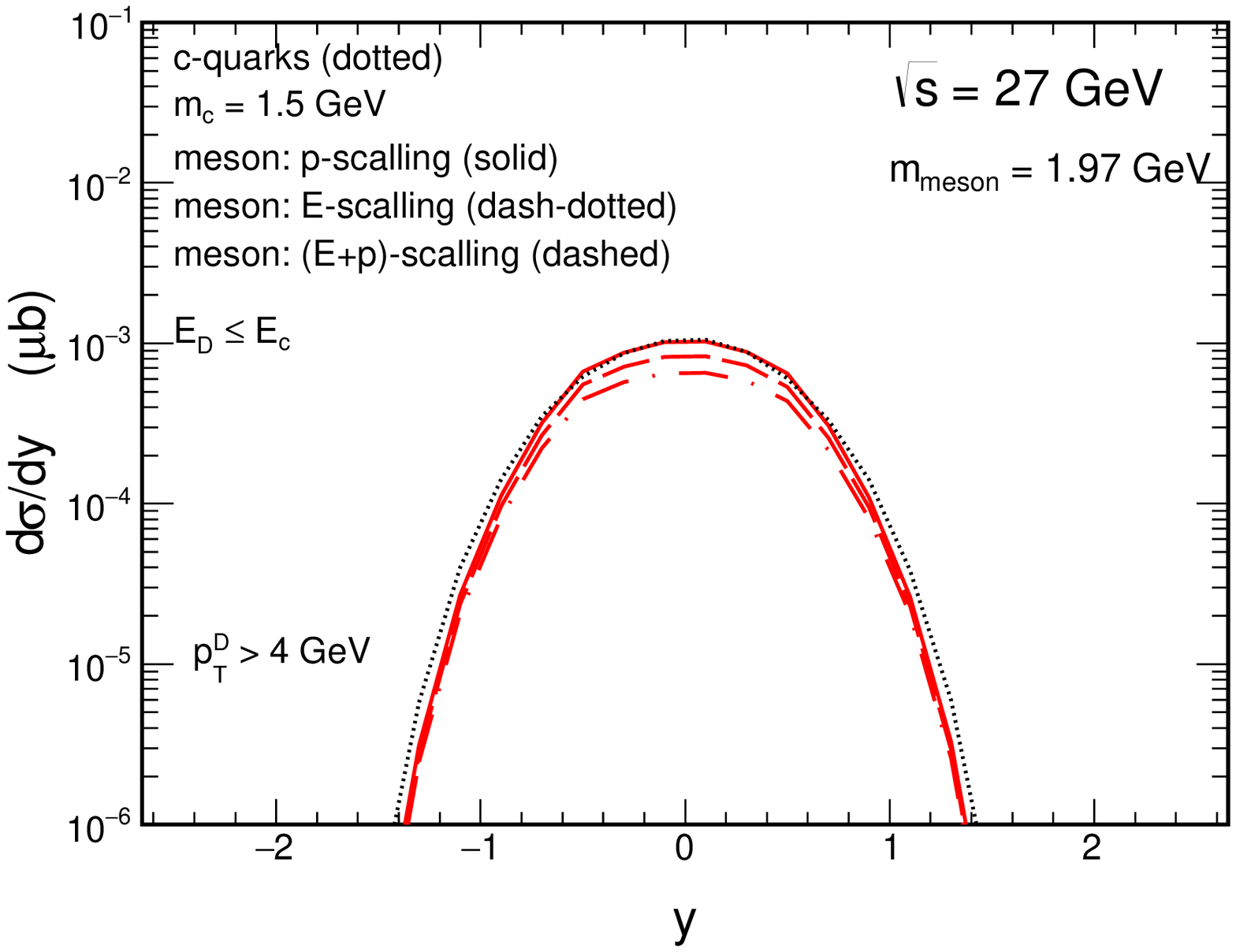}}
\end{minipage}\\
\begin{minipage}{0.47\textwidth}
  \centerline{\includegraphics[width=1.0\textwidth]{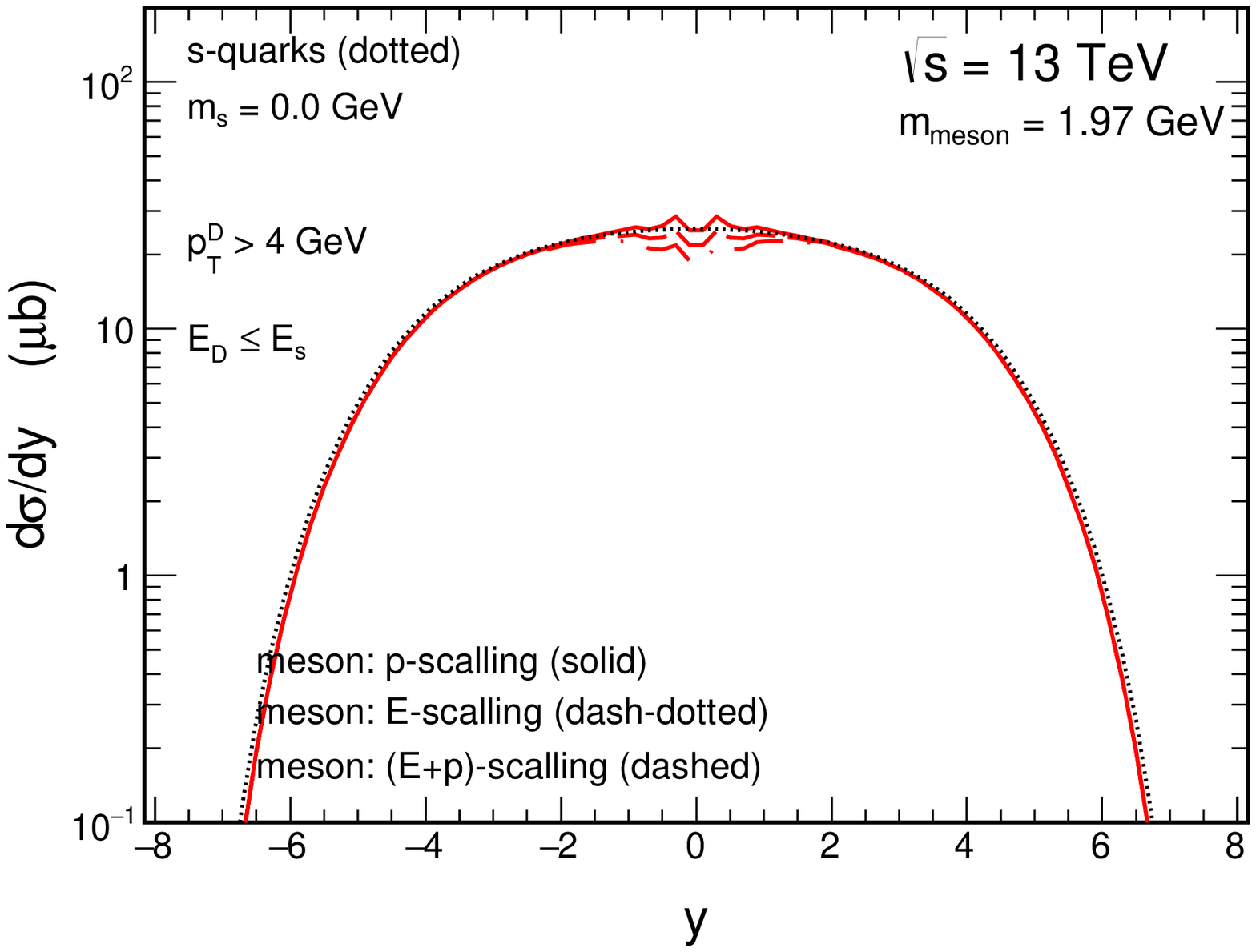}}
\end{minipage}
\begin{minipage}{0.47\textwidth}
  \centerline{\includegraphics[width=1.0\textwidth]{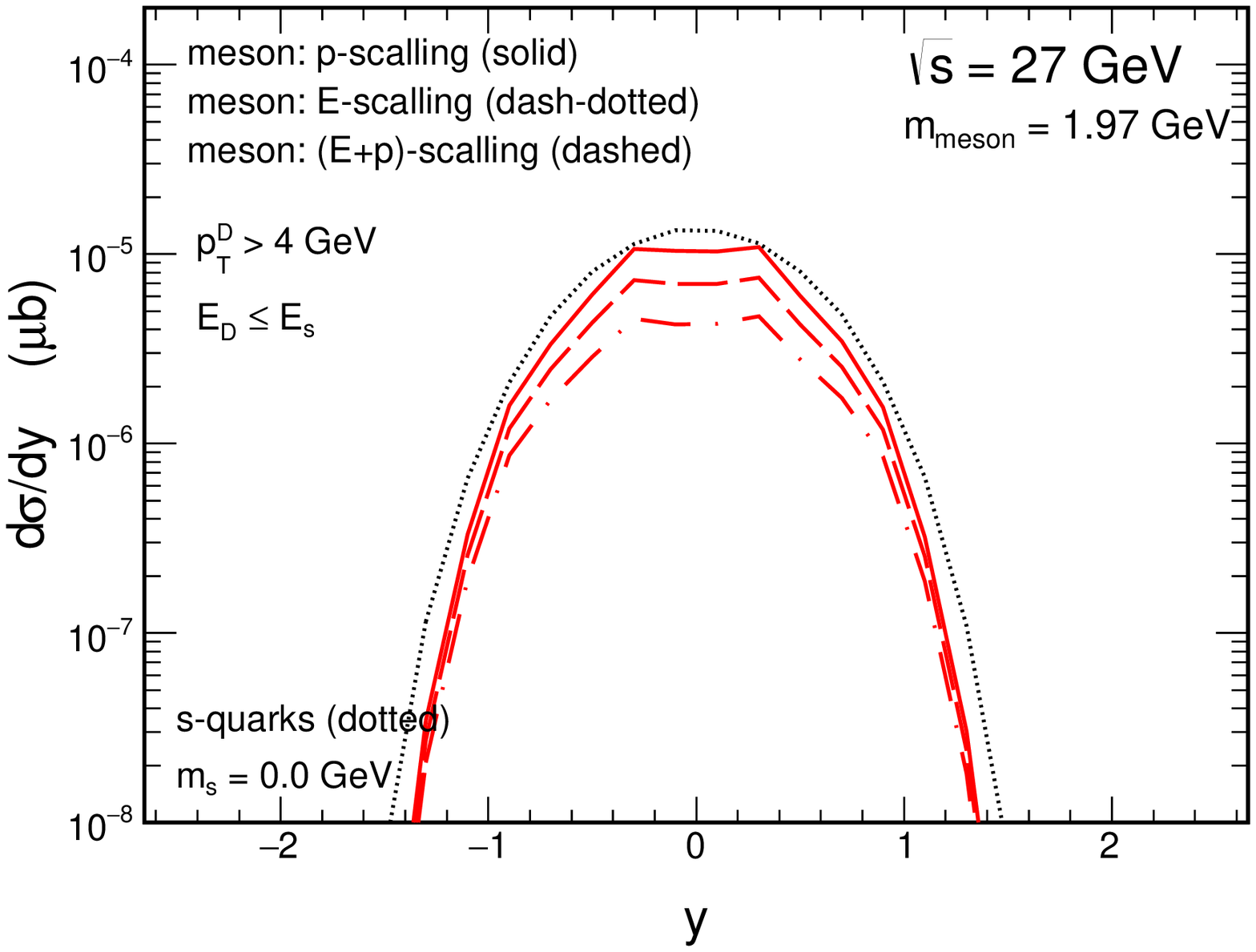}}
\end{minipage}
\caption{\small
Rapidity distributions of $D_s^{\pm}$ mesons for $c/\bar c \to
D_s^{\pm}$ (top panels) and $s/\bar s \to D_s^{\mp}$ (bottom panels)
for $\sqrt{s}$ = 13 TeV (left) and $\sqrt{s}$ = 27 GeV (right).
Here $p_{t,D_s} >$ 4 GeV.
}
\label{fig:dsig_dy_ptDgt4}
\end{figure}

In Figs.~\ref{fig:dsig_dy_ptDgt2} and ~\ref{fig:dsig_dy_ptDgt4} we show rapidity distributions
for two lower cuts on transverse momentum of $D_s$ meson, $p_{t} > 2$ and $4$ GeV, respectively.
For low transverse momentum cuts the subleading contribution may be
as important as the leading one. Above $p_{t} =$ 4 GeV the contribution of the subleading fragmentation
is, however, much smaller than the leading one and can be safely neglected.
In this region, at high energy, the results for different fragmentation methods start to coincide and
one recovers the standard approach. This is not true in the case of low energy, where still some differences remain noticeable,
for both, $c$- and $s$-quark fragmentation.
  
\begin{figure}[!h]
\begin{minipage}{0.47\textwidth}
  \centerline{\includegraphics[width=1.0\textwidth]{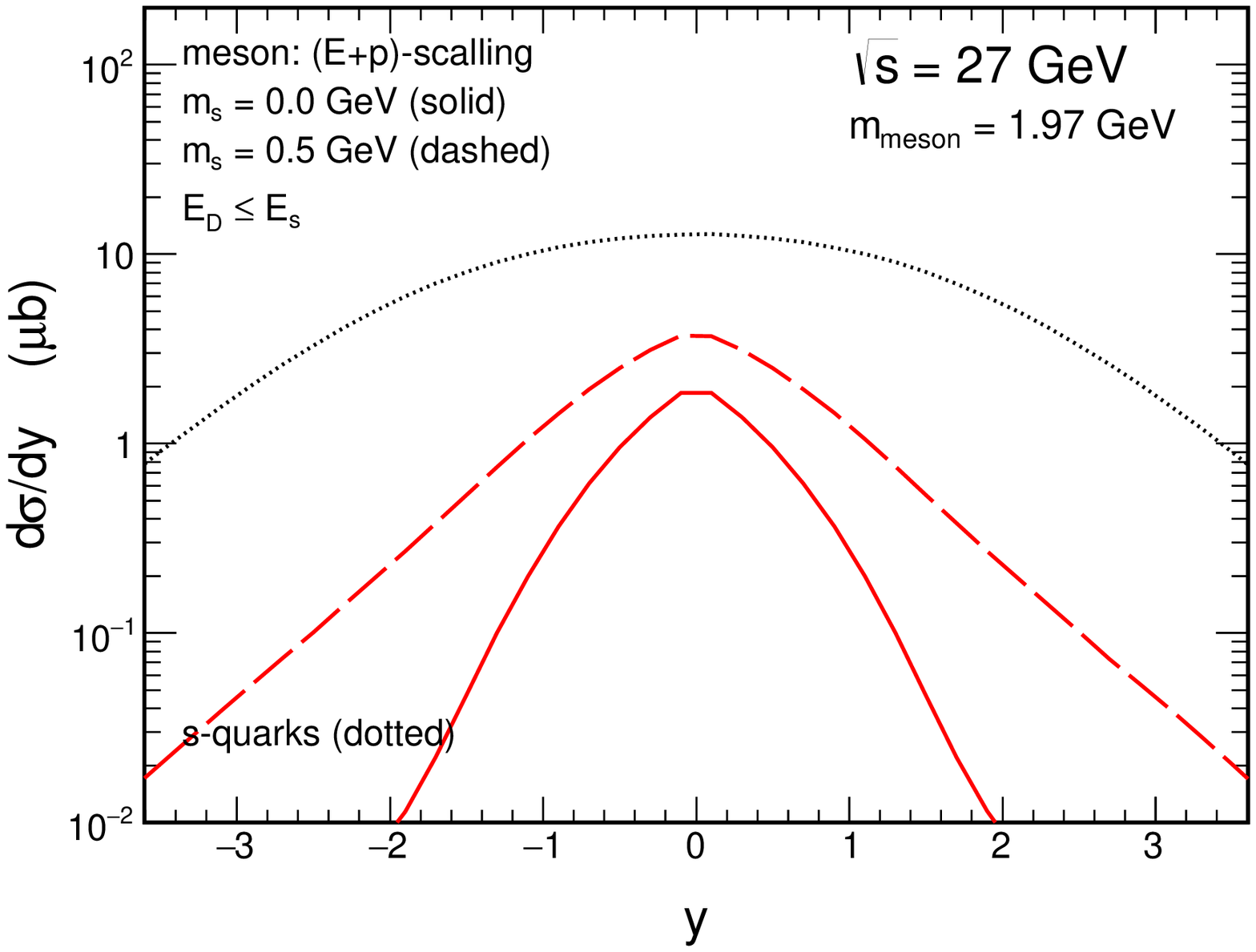}}
\end{minipage}
\begin{minipage}{0.47\textwidth}
  \centerline{\includegraphics[width=1.0\textwidth]{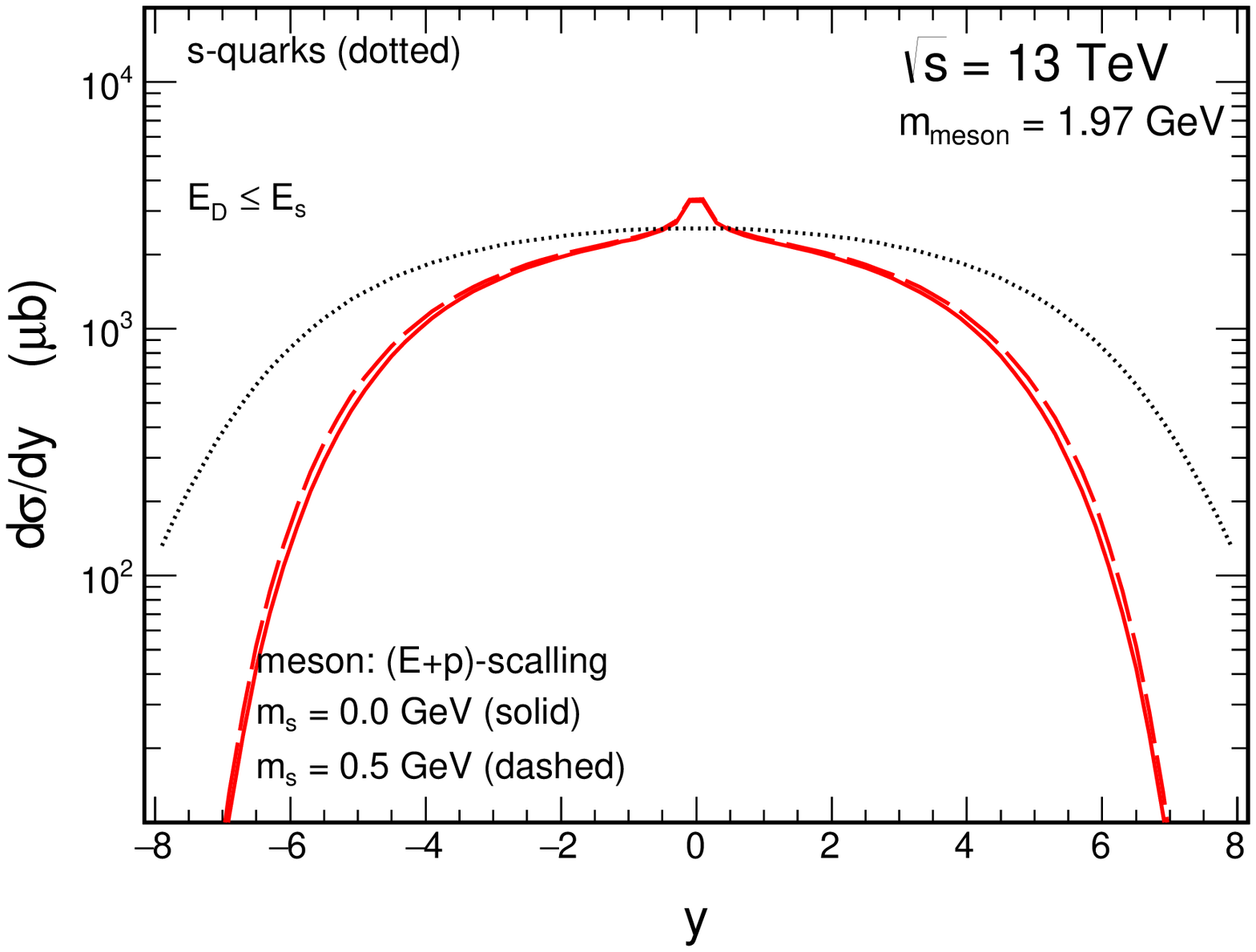}}
\end{minipage}
\caption{\small
Rapidity distributions of $D_s^{\pm}$ mesons from 
$s/{\bar s} \to D_s^{\pm}$ fragmentation for $\sqrt{s}=27$ GeV (left) and $\sqrt{s}=13$ TeV (right)
for massless (solid) and for massive (dash-dotted) treatment of $s$-quarks in the fragmentation procedure.
}
\label{fig:dsig_dy_light_to_Ds_smass}
\end{figure}

Let us now discuss a possible role of the $s$-quark mass in the 
light-to-heavy fragmentation procedure.
In Fig.~\ref{fig:dsig_dy_light_to_Ds_smass} we show rapidity
distributions for $s$-quarks and $D_s$ mesons from their fragmentation calculated with momentum scaling method for the two considered energies (left and right panels) for two cases: taking $m_s =$ 0 GeV (solid lines) and $m_s =$ 0.5 GeV (dash-dotted lines)
in the fragmentation procedure. The mass effects are only significant at low energy and in the region of small meson transverse momenta. At high energy
and/or with lower $p_t$-cut it will become completely negligible.   

\begin{figure}[!h]
\begin{minipage}{0.47\textwidth}
  \centerline{\includegraphics[width=1.0\textwidth]{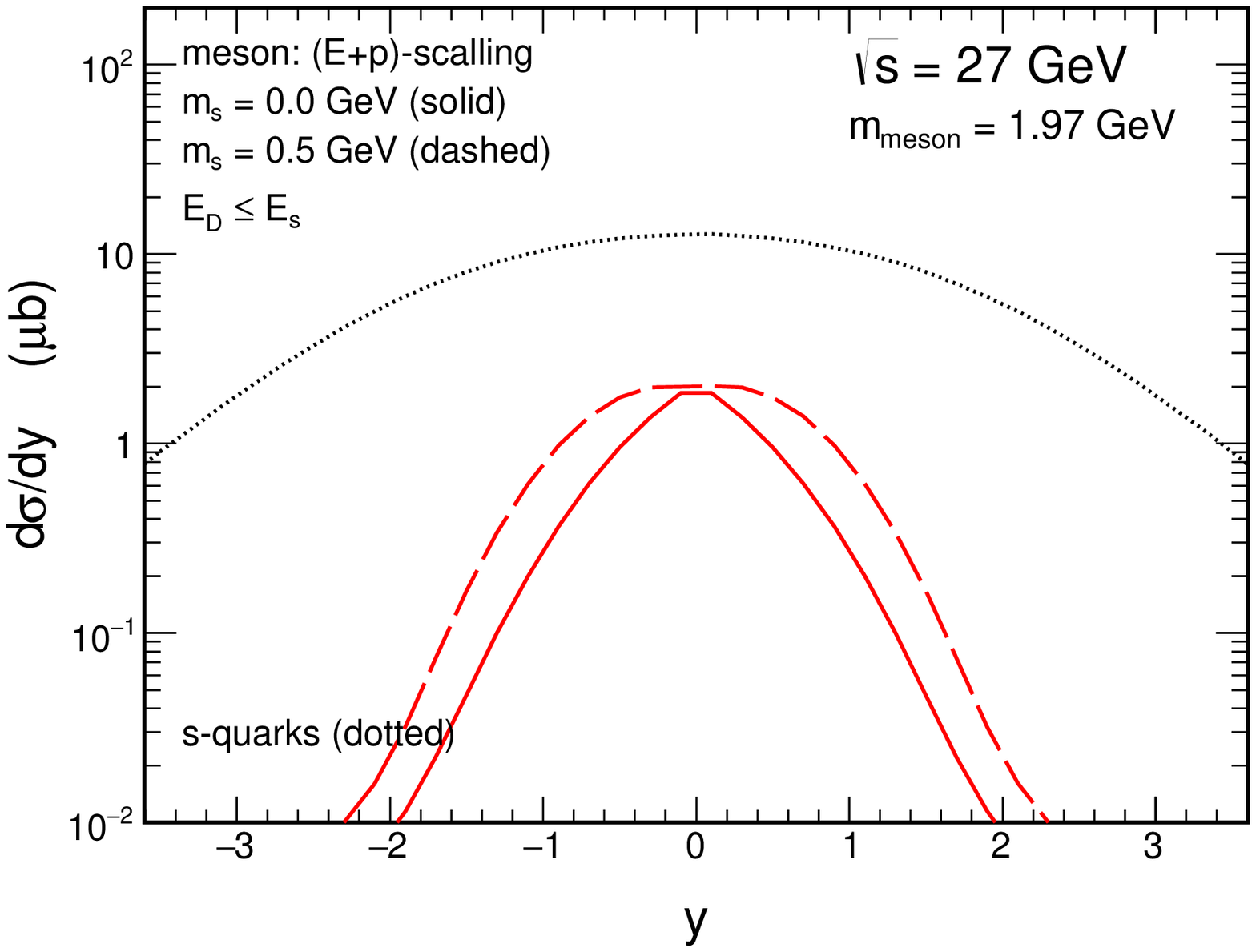}}
\end{minipage}
\begin{minipage}{0.47\textwidth}
  \centerline{\includegraphics[width=1.0\textwidth]{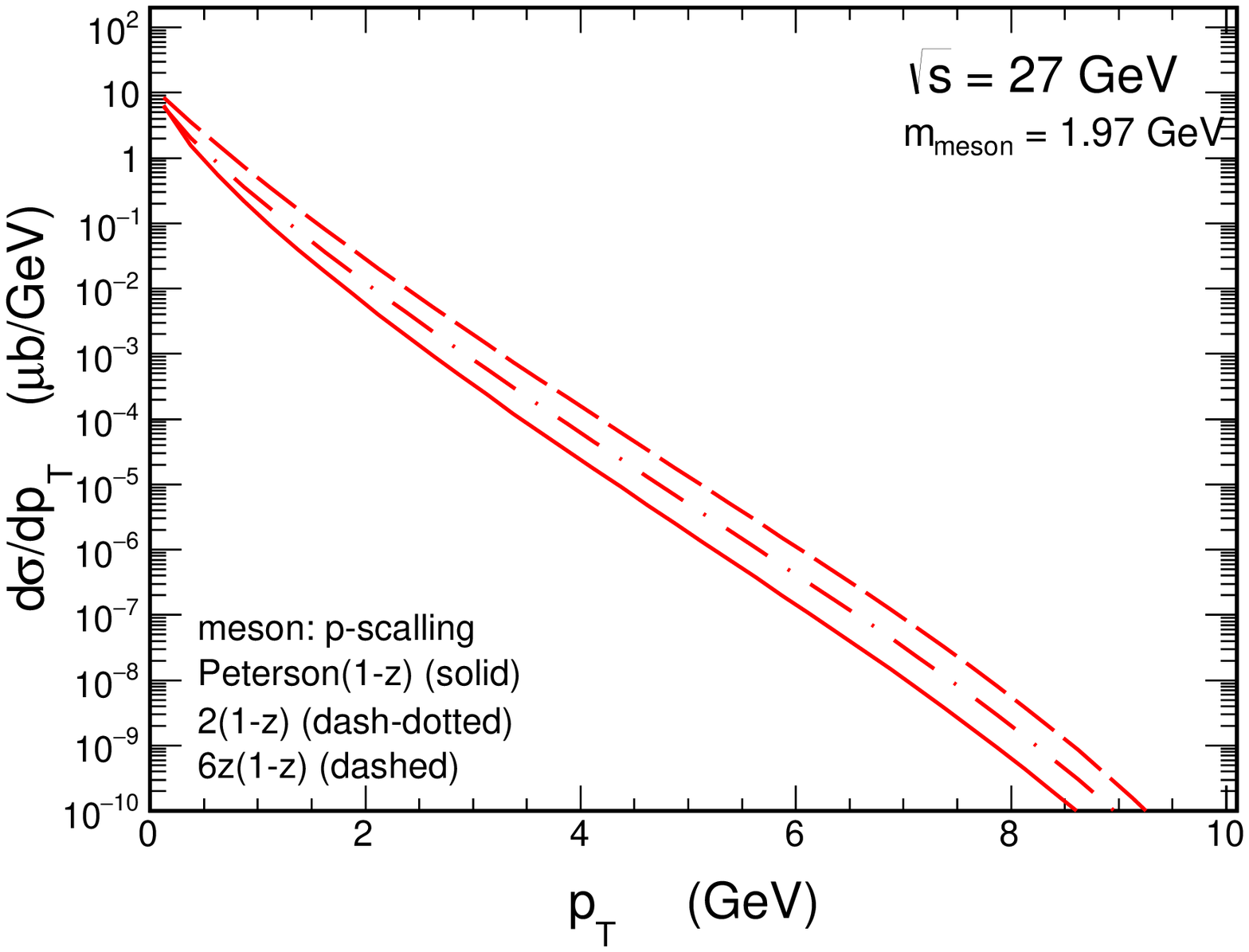}}
\end{minipage}
\begin{minipage}{0.47\textwidth}
  \centerline{\includegraphics[width=1.0\textwidth]{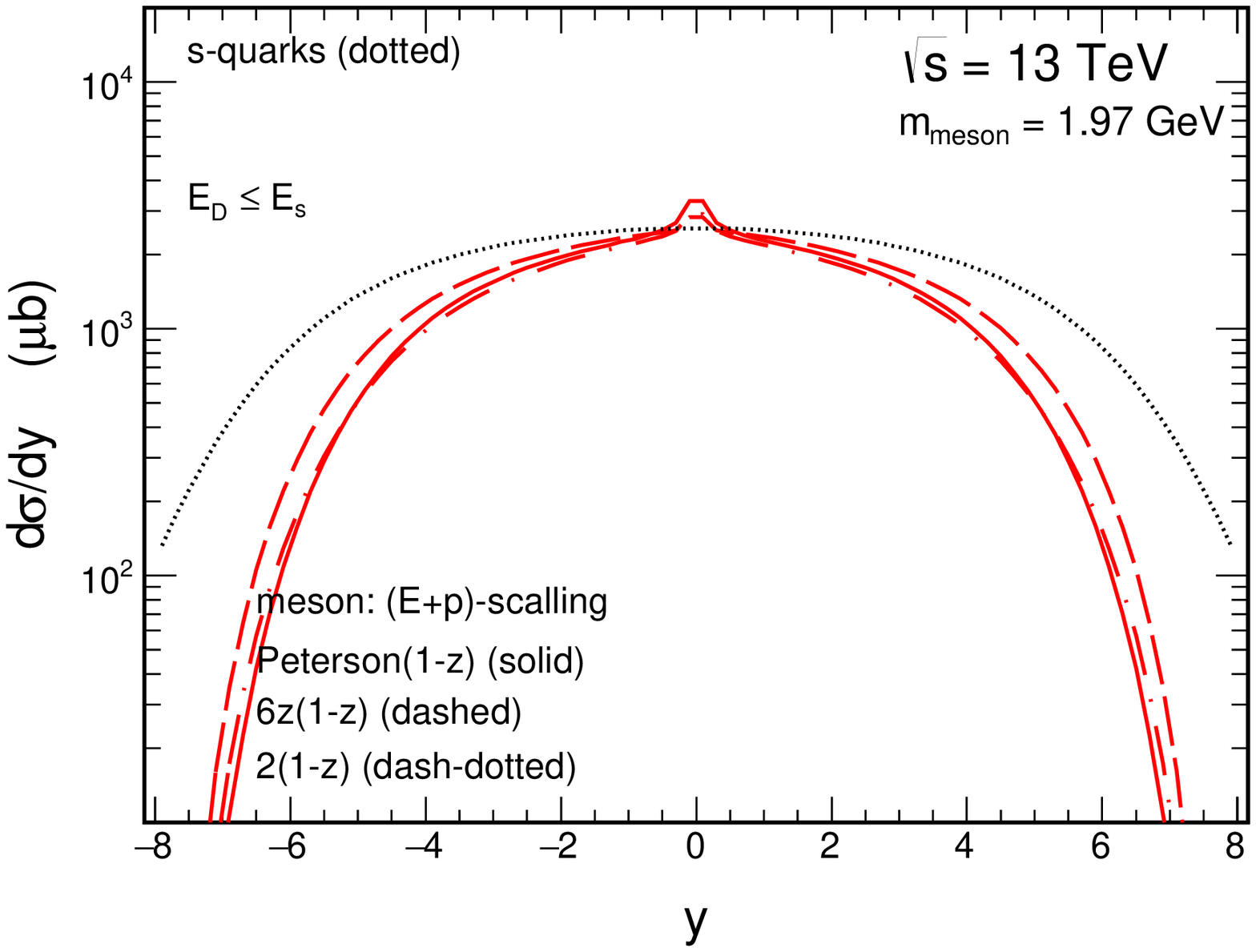}}
\end{minipage}
\begin{minipage}{0.47\textwidth}
  \centerline{\includegraphics[width=1.0\textwidth]{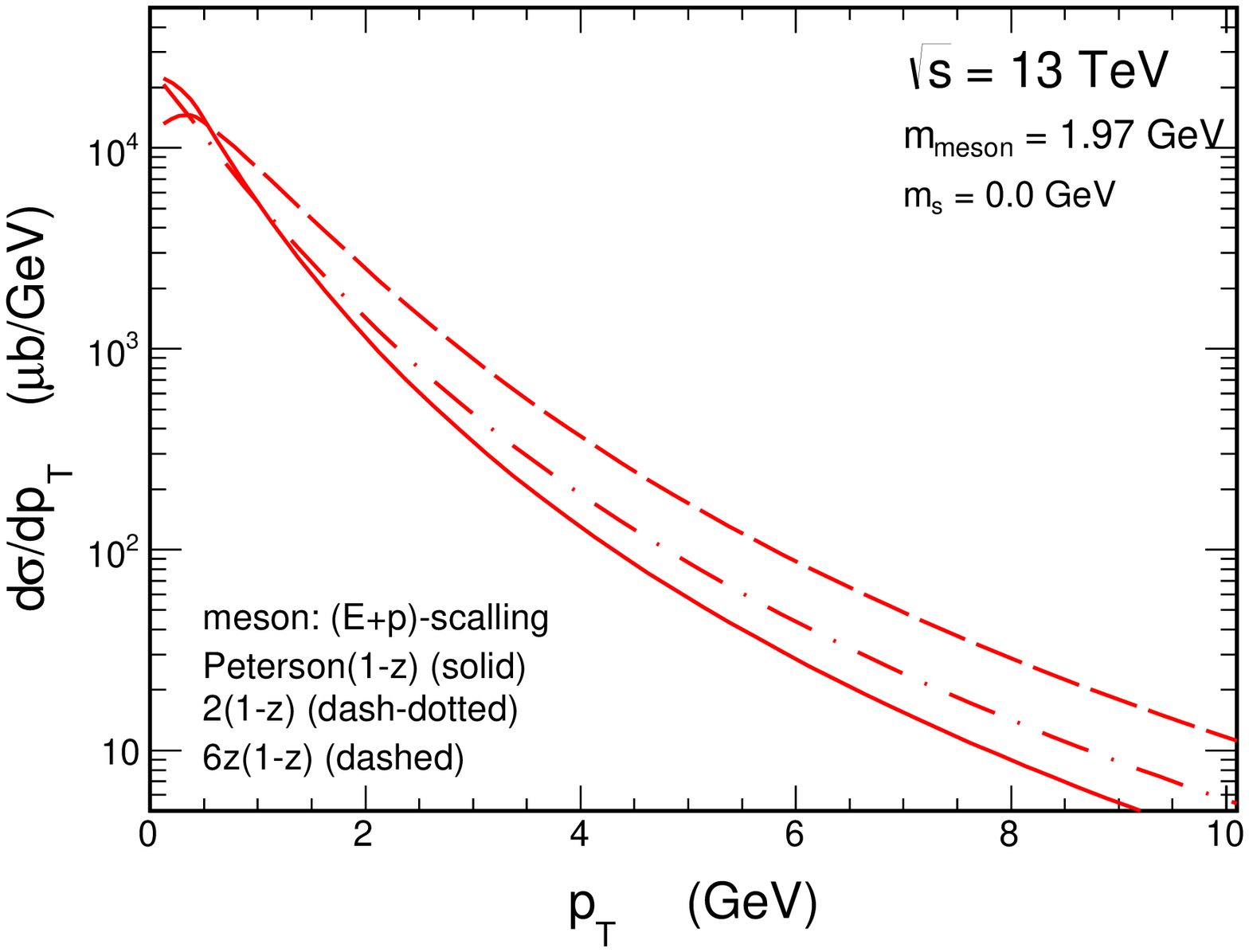}}
\end{minipage}
\caption{\small
Rapidity (left) and transverse momentum (right) distributions of $D_s^{\pm}$ mesons from 
$s/{\bar s} \to D_s^{\pm}$ fragmentation for $\sqrt{s}=27$ GeV (top) and $\sqrt{s}=13$ TeV (bottom)
for different parametrizations of fragmentation functions.
}
\label{fig:different_FF}
\end{figure}

Finally, in Fig.~\ref{fig:different_FF} we discuss effects related to the choice of the different parametrizations
for $s / {\bar s} \to D_s^{\mp}$ fragmentation function. Here we use momentum scaling method with different parametrizations of fragmentation functions.
We show a visible sensitivity of our results to the choice of the FFs, that may become really large when going to large meson transverse momenta.
The effect for small transverse momenta of $D_s$ mesons is rather small.

\section{Conclusions}

We have made a critical analysis of independent parton fragmentation 
of quarks/antiquarks to heavy mesons in proton-proton collisions. 
We have shown that different approaches in the literature lead 
to different results.
We have considered both light quark/antiquark fragmentation
(called here light-to-heavy) and heavy quark/antiquark fragmentation
(called here heavy-to-heavy).

A special emphasis has been made for fragmentation along the direction
of parton, not discussed so far in the literature on the subject.
We have compared results obtained for scaling in different variables
(momentum, energy, light-cone).
As an example we have considered production of $D_s$ mesons.
The latter are very important e.g. for production of $\nu_{\tau}$
or ${\bar \nu}_{\tau}$ discussed in the context of IceCube 
\cite{Aartsen:2015dlt} or SHiP \cite{Bai:2018xum} experiments and 
asymmetry of $D_s^{\pm}$ mesons \cite{Goncalves:2018zzf}
observed recently by the LHCb \cite{Aaij:2018afd}.

Different results have been obtained for different approaches,
especially for light-to-heavy fragmentation.
It has been discussed that some approaches lead to energy violation
and other approaches to flavour violation which in a strong process,
such as fragmentation, should be conserved.
The ways out have been suggested.

As an example we have also shown rapidity distributions of $D_s$ mesons.
Both heavy-to-heavy and light-to-heavy contributions have been compared.
The presence of the subleading contribution may potentially influence
the extraction of $P(c / \bar c \to D_s)$ made so far in the literature.

As we have illustrated in the present paper, the effect of using
different prescriptions is particularly large at low energies,
e.g. fixed target experiments.
Therefore the considerations presented in this paper are very important 
for simulations of $\nu_{\tau}/\bar{\nu_{\tau}}$ production for fixed 
target experiments such as SHIP \cite{SHIP}. This will be discussed 
elsewhere \cite{MSZ2019}.



\end{document}